\documentclass[twocolumn]{aastex63}

\usepackage{apjfonts}
\usepackage{color}
\usepackage{soul}
\usepackage{natbib}
\usepackage{hyperref}
\usepackage{amsmath}
\usepackage{xspace}
\usepackage{graphicx}
\usepackage{enumitem}
\usepackage{amssymb}
\usepackage{xifthen}
\usepackage[normalem]{ulem}
\usepackage{aas_macros}
\usepackage{amsfonts}
\usepackage{amsmath}
\usepackage{multirow}
\usepackage{dsfont}
\usepackage{hyperref}

\newcommand{\angstrom}{\mbox{\normalfont\AA}\xspace}
\newcommand{\SSSSS}{${S}^5$\xspace}
\newcommand{\code}[1]{\texttt{#1}\xspace}

\hypersetup{    
  colorlinks      = {true},
  linkcolor       = {blue},
  citecolor       = {blue},
  urlcolor        = {blue},
}

\shortauthors{Webber et al.}

\begin{document}
\title{Detailed Chemical Abundance Analysis of the Brightest Stars in the Turranburra and Willka Yaku Stellar Streams

\footnote{This paper includes data gathered with the 6.5m Magellan Telescopes located at Las Campanas Observatory, Chile.}}

\author[0000-0002-9762-4308]{K.~B.~Webber}
\affil{Mitchell Institute for Fundamental Physics and Astronomy and Department of Physics and Astronomy, Texas A\&M University, College Station, TX 77843-4242, USA}

\author[0000-0001-6154-8983]{T.~T.~Hansen}
\affil{Department of Astronomy, Stockholm University, AlbaNova
University Center, SE-106 91 Stockholm, Sweden}

\author{J.~L.~Marshall}
\affil{Mitchell Institute for Fundamental Physics and Astronomy and Department of Physics and Astronomy, Texas A\&M University, College Station, TX 77843-4242, USA}

\author[0000-0002-4863-8842]{Alexander~P.~Ji}
\affil{Department of Astronomy \& Astrophysics, University of Chicago, 5640 S Ellis Avenue, Chicago, IL 60637, USA}
\affil{Kavli Institute for Cosmological Physics, University of Chicago, Chicago, IL 60637, USA}
\affil{NSF-Simons AI Institute for the Sky (SkAI), 172 E. Chestnut St., Chicago, IL 60611, USA}

\author[0000-0002-9110-6163]{Ting~S.~Li}
\affil{Department of Astronomy and Astrophysics, University of Toronto, 50 St. George Street, Toronto ON, M5S 3H4, Canada}
\affil{Dunlap Institute for Astronomy \& Astrophysics, University of Toronto, 50 St George Street, Toronto, ON M5S 3H4, Canada}
\affil{Data Sciences Institute, University of Toronto, 17th Floor, Ontario Power Building, 700 University Ave, Toronto, ON M5G 1Z5, Canada}

\author[0000-0001-7019-649X]{Gary~S.~Da~Costa}
\affil{Research School of Astronomy and Astrophysics, Australian National University, Canberra, ACT 2611, Australia}
\author[0000-0001-8536-0547]{Lara~R.~Cullinane}
\affil{Leibniz-Institut f{\"u}r Astrophysik Potsdam (AIP), An der Sternwarte 16, D-14482 Potsdam, Germany}

\author[0000-0002-8448-5505]{Denis~Erkal}
\affil{Department of Physics, University of Surrey, Guildford GU2 7XH, UK}

\author[0000-0003-2644-135X]{Sergey~E.~Koposov}
\affil{Institute for Astronomy, University of Edinburgh, Royal Observatory, Blackford Hill, Edinburgh EH9 3HJ, UK}
\affil{Institute of Astronomy, University of Cambridge, Madingley Road, Cambridge CB3 0HA, UK}
\author[0000-0003-0120-0808]{Kyler~Kuehn}
\affil{Lowell Observatory, 1400 W Mars Hill Rd, Flagstaff,  AZ 86001, USA}

\author[0000-0003-3081-9319]{Geraint~F.~Lewis}
\affil{Sydney Institute for Astronomy, School of Physics, A28, The University of Sydney, NSW 2006, Australia}
\author[0000-0002-6529-8093]{Dougal~Mackey}
\affil{Research School of Astronomy and Astrophysics, Australian National University, Canberra, ACT 2611, Australia}
\author[0000-0002-3430-4163]{Sarah~L.~Martell}
\affil{School of Physics, University of New South Wales, Sydney, NSW 2052, Australia}
\author[0000-0002-6021-8760]{Andrew~B.~Pace}
\affil{Department of Astronomy, University of Virginia, 530 McCormick Road, Charlottesville, VA 22904, USA}
\author[0000-0003-2497-091X]{Nora~Shipp}
\affil{Department of Astronomy, University of Washington, Seattle, WA 98195, USA}
\author[0000-0002-8165-2507]{Jeffrey~D.~Simpson}
\affil{School of Physics, University of New South Wales, Sydney, NSW 2052, Australia}
\author[0000-0002-3105-3821]{Zhen~Wan}
\affil{Sydney Institute for Astronomy, School of Physics, A28, The University of Sydney, NSW 2006, Australia}
\author[0000-0003-1124-8477]{Daniel~B.~Zucker}
\affil{School of Mathematical and Physical Sciences, Macquarie University, Sydney, NSW 2109, Australia}
\affil{Macquarie University Research Centre for Astrophysics and Space Technologies, Sydney, NSW 2109, Australia}
\author{Victor~A.~Alvarado}
\affil{Mitchell Institute for Fundamental Physics and Astronomy and Department of Physics and Astronomy, Texas A\&M University, College Station, TX 77843-4242, USA}

\author[0000-0001-7516-4016]{Joss~Bland-Hawthorn}
\affil{Sydney Institute for Astronomy, School of Physics, A28, The University of Sydney, NSW 2006, Australia}
\affil{Centre of Excellence for All-Sky Astrophysics in Three Dimensions (ASTRO 3D), Australia}
\author[0000-0002-9269-8287]{Guilherme~Limberg}
\affil{Kavli Institute for Cosmological Physics, University of Chicago, Chicago, IL 60637, USA}
\author[0000-0003-0105-9576]{Gustavo~E.~Medina}
\affil{Department of Astronomy and Astrophysics, University of Toronto, 50 St. George Street, Toronto ON, M5S 3H4, Canada}
\affil{Dunlap Institute for Astronomy \& Astrophysics, University of Toronto, 50 St George Street, Toronto, ON M5S 3H4, Canada}
\author[0000-0003-0918-7185]{Sam~A.~Usman}
\affil{Department of Astronomy \& Astrophysics, University of Chicago, 5640 S Ellis Avenue, Chicago, IL 60637, USA}
\affil{Kavli Institute for Cosmological Physics, University of Chicago, Chicago, IL 60637, USA}

\correspondingauthor{K.~B.~Webber}
\email{kbwebber@tamu.edu}

\begin{abstract}
 We present a detailed chemical abundance analysis of the three brightest known stars from each of the Turranburra and Willka Yaku stellar streams using high-resolution Magellan/MIKE spectra. Abundances for 27 elements, ranging from carbon to dysprosium, were derived. Our results support the original classification that Turranburra, with a low average metallicity of $\mathrm{[Fe/H]=-2.45} \pm 0.07$, likely originates from a dwarf-galaxy progenitor. Willka Yaku has a low average metallicity of $\mathrm{[Fe/H]=-2.35 \pm 0.03}$ with a small scatter in the abundances, consistent with a globular cluster progenitor as suggested by previous studies. Both streams exhibit mild enhancements in neutron-capture elements, with averages of $\mathrm{[Eu II/Fe]}=$ $0.47 \pm{0.09}$ for Turranburra and $0.44 \pm{0.05}$ for Willka Yaku, consistent with enrichment from an $r$-process event. A similar enrichment is observed in other stellar streams, and we further discuss this signature as it relates to the potential enrichment histories of these two streams.
\end{abstract}

\section{Introduction}

The Milky Way's (MW) stellar halo is composed of remnants from past mergers and accretion events \citep{johnston2008, helmi2020, bonaca2025}. Stellar streams represent the first stage of this accretion process, where stellar associations such as dwarf galaxies (DGs) and globular clusters (GCs) have been tidally disrupted by the MW but remain spatially and kinematically distinct. These structures can be detected via large-scale photometric surveys such as the Dark Energy Survey \citep[DES;][]{DES, shipp2018}. At later stages, stars from these disrupted systems become fully phase-mixed with the original MW stars but retain their infall angular momentum. Thus, dynamical groups of stars accreted from the same parent object can be identified through their kinematics using $Gaia$ data \citep{gaia2016}. 

The Southern Stellar Stream Spectroscopic Survey (\SSSSS; \citealt{li2019}) combines DES data with $Gaia$ parallaxes and proper motions to determine stream dynamics and identify candidate stream members. These candidate members are then observed spectroscopically to obtain a radial velocity and confirm or reject membership status. In addition to the work of \SSSSS, many streams have been discovered in the $Gaia$ dataset using the {\tt STREAMFINDER} algorithm, which implements a probabilistic approach to identify stellar structures by analyzing their orbital dynamics and phase-space properties \citep{ibata2018}.

Most known stellar streams originating from DG infall are found to be remnants of classical DGs ($10^6 \lesssim M_*/M_\odot \le 10^9$), such as the Wukong and Indus stellar streams \citep{limberg2024, hansen2021, ji2020a}. However, models of galaxy formation suggest that there should be many more low-mass stellar streams ($\lesssim 10^5$ M$_\odot$) in the MW halo, originating from ultra-faint dwarf (UFD) galaxies \citep{johnston1998, shipp2023}. Finding such systems would significantly add to our understanding of the build-up of the MW by, for example, showing how low-mass substructures have contributed to its dark-matter assembly. 

The first example of a UFD galaxy stream might be the Leiptr stellar stream, which is predicted to have come from a progenitor with mass $\lesssim 10^5 $ M$_\odot$ \citep{atzberger2024}. Since stellar streams retain the chemical signature of their progenitor, chemical analysis can tell us about the nature of the progenitor. The chemical analysis of the Leiptr stream found very low abundances of neutron-capture elements ($\mathrm{[Eu/Fe]} \sim -0.4$), supporting the UFD progenitor scenario, as most intact UFD galaxies display deficiencies in neutron-capture elements \citep{ji2019a}. 

However, not all UFD galaxies show low neutron-capture element abundances. There are currently four known UFD galaxies: Reticulum~II, Tucana~III, Grus~II, and Tucana~V, that contain stars enriched in rapid neutron-capture process ($r$-process) elements, $\mathrm{[Eu/Fe]\gtrsim 0.3}$ \citep{christlieb2004, beers2005, holmbeck2020}. The Reticulum~II UFD galaxy is the most $r$-process enhanced system with $72\%$ of the stars showing an enhancement \citep{ji2016c, ji2023a}, while four of the five stars analyzed in the Tucana~III UFD galaxy show an average enhancement of $\mathrm{[Eu/Fe]} \sim 0.4$  \citep{marshall2019}. Finally, the Grus~II and Tucana~V galaxies have one star each that have $\mathrm{[Eu/Fe]}$ abundances of $0.31$ and $0.36$ respectively \citep{hansen2017, hansen2020, hansen2024}. Stellar streams originating from UFD galaxies like these four, would also show an $r$-process enriched chemical signature.

In this paper, we present a chemical analysis of three stars in each of the Turranburra and Willka Yaku stellar streams. Both streams were discovered in DES data \citep{shipp2018} and later spectroscopically followed-up on in the \SSSSS collaboration. Turranburra has a length of 8.1 kpc, width of 288 pc, and a stellar mass of $7.6\times10^3$M$_\odot$ while Willka Yaku has a length of 3.9 kpc, width of 127 pc, and stellar mass of $4.6\times10^3$M$_\odot$ \citep{shipp2018}. \cite{li2022} measured radial velocities for 22 and nine members in Turranburra and Willka Yaku respectively. The outline of this paper is as follows. In Section \ref{sec:observations}, the observations are described, and Section \ref{sec:stellar parameters and abundance analysis} describes the stellar parameters and analysis. Section \ref{sec:results} presents the results, which are further discussed in Section \ref{sec:discussion}, and Section \ref{sec:summary} provides a summary. 

\section{Observations \label{sec:observations}}
\begin{figure*}
    \centering
    \includegraphics[scale=0.5]{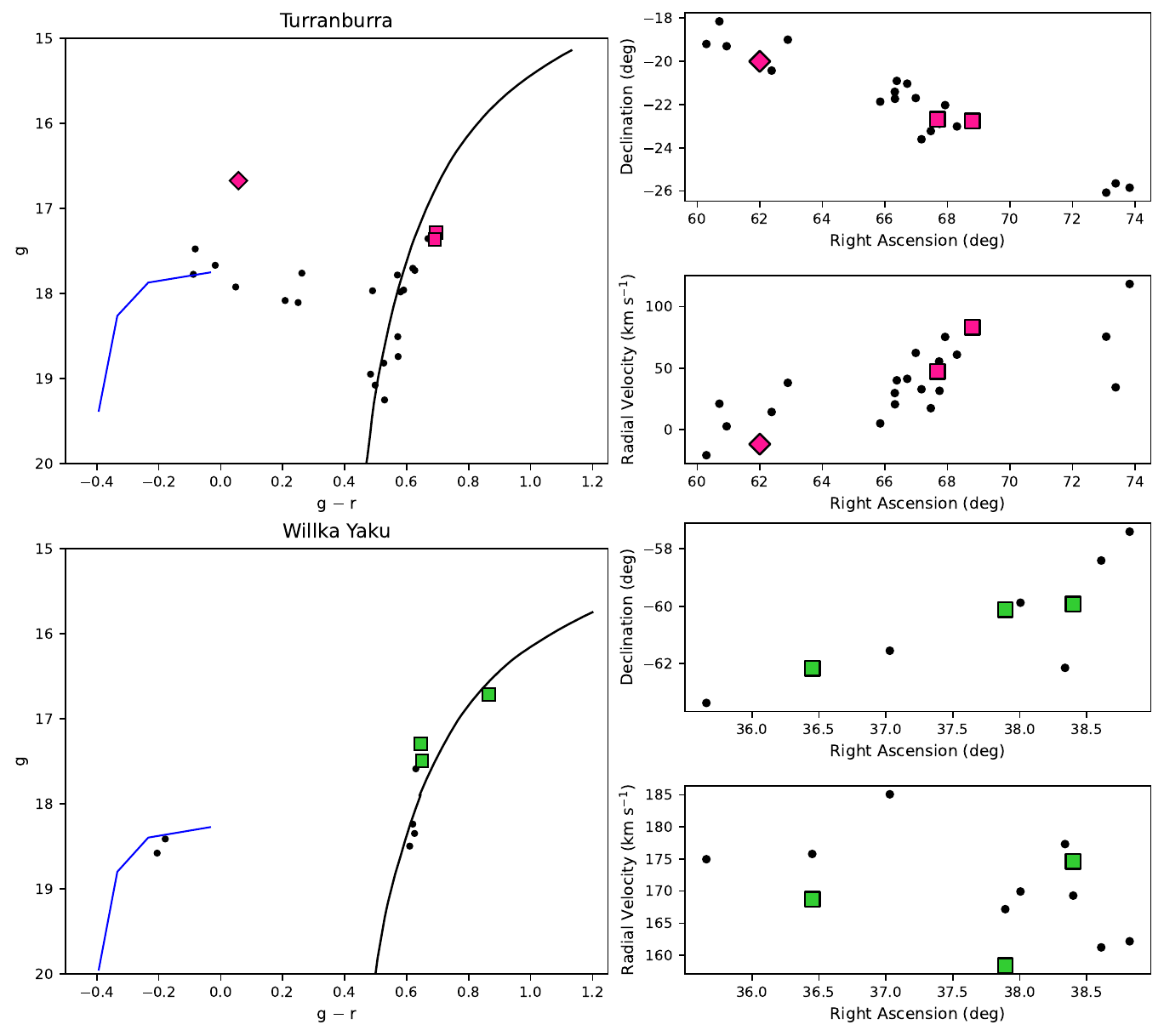}
    \caption{\label{Fig:cmd}The left panels show the color-magnitude diagrams for the Turranburra and Willka Yaku fields. The right panels show the Declination vs. Right Ascension (Top) and Radial Velocity vs Right Ascension (Bottom) for the stream stars. Both streams were first reported in \cite{shipp2019} and further characterized by \cite{li2022}; we refer the interested reader to those papers for further details on the membership selection. The pink squares and diamond are the Turranburra stars presented in this paper, and the diamond represents Tur-3, an RRL. The green squares are the Willka Yaku stars presented in this paper. For both streams, the black circles are other member stars from \cite{li2022}. MIST isochrones \citep{dotter2016} shown as black lines were produced using the DECam photometric system, using an age of 12.5 Gyr for both streams, with metallicities taken from \cite{shipp2019}, of Z= 0.0006 and 0.0003, respectively, and shifted to the distances of Turranburra and Willka Yaku (27.5 kpc and 34.7kpc, respectively). The blue lines represent the horizontal branch isochrone from M92.} 
\end{figure*}

Initial spectroscopic observations of the systems were obtained by the \SSSSS collaboration using the Two-Degree Field (2dF) fiber positioner \citep{lewis2002} with the AAOmega spectrograph \citep{sharp2006} on the Anglo-Australian Telescope (AAT) to measure radial velocities and metallicities of member stars and further characterize the systems \citep{li2022}. These initial observations indicated that Turranburra likely comes from a DG progenitor, while Willka Yaku is likely from a GC progenitor. As further detailed by \cite{li2022}, members of the Turranburra and Willka Yaku streams were selected using kinematic cuts, and for Turranburra, a metallicity cut ($\mathrm{[Fe/H]<-1.5}$) was also applied to avoid foreground stars (see \cite{li2022} for details). High-resolution spectra was subsequently taken in October 2021 with the Magellan Inamori Kyocera Echelle (MIKE) spectrograph \citep{bernstein2003} of the three brightest member stars identified by \cite{li2022} in Turranburra and Willka Yaku.
 The spectra were obtained using the $0\farcs7$ slit and 2$\times$2 pixel binning, covering a wavelength range of 3350-5000~\angstrom in the blue and 4900-9500~\angstrom in the red, with resolving powers $R = \lambda/\Delta\lambda \sim$35,000 at $\lambda\approx4200$~\angstrom and $\sim$28,000 at $\lambda\approx7200$~\angstrom, respectively. The data were reduced using the CarPy MIKE pipeline \citep{kelson2000, kelson2003}. Following reduction, the spectra were normalized and shifted to rest wavelength. Color-magnitude diagrams for Turranburra and Willka Yaku member stars \citep{li2022} are shown in Figure \ref{Fig:cmd} where the stars analyzed in this paper are marked with pink and green squares respectively. Table \ref{tab:obs} presents the details of the observations, namely their $Gaia$ DR3 source\_id, coordinates, heliocentric Julian date (HJD) of the MIKE observations, exposure times, $Gaia$ $G$, $BP$, and $RP$ photometry, signal-to-noise ratios (SNR), and heliocentric radial velocities. The heliocentric radial velocities were determined by cross-correlation of each order with a spectrum of HD~122563 ($V_{\rm helio}$ = $-$26.13 $\pm 0.04$ km~s$^{-1}$; \citealt{gaiarv21}) obtained with the same setup. The standard deviation of the radial velocities from the orders used in the cross-correlation is reported as the uncertainty. The number of orders used for the cross-correlation (N$_{orders}$) is also listed in Table \ref{tab:obs}. The velocities were compared to the values reported by \cite{li2022} which were $48.7 \pm{0.5}$km~s$^{-1}$, and $84.1\pm{0.9}$km~s$^{-1}$ for Tur-1 and 2 respectively, and Tur-3 was not reported in \cite{li2022}. The velocities for WY-1, 2, and 3 are $167.2 \pm{0.3}$km~s$^{-1}$, $175.8 \pm{0.6}$km~s$^{-1}$, and $169.3 \pm{0.5}$ respectively. All of the velocities from \cite{li2022} are in good agreement with those reported here, except for WY-1, which shows a velocity variation that could suggest it is in a binary system.

\begin{deluxetable*}{lccccccccccc}
\caption{Observing Details and Stellar Data \label{tab:obs}}
\tablehead{Name & Source ID & RA&  DEC & HJD & $t_{exp}$ & $G$ & $BP$ & $RP$ &  $SNR$  & $V_{\rm helio}$ & N$_{orders}$ \  \\& ($Gaia$ DR3) & & & &(s) &(mag) &(mag) & (mag) &(@4800 \AA)&(km~s$^{-1}$)&  }
\startdata
Tur-1 & 4898141752749430400 & 04:30:43.67 & -22:40:36.5 &  2459513.76957 & 4x1800 & 16.64 & 17.13 & 15.98 & 28  & 47.4 $\pm$ 1.3 & 51 \\
Tur-2 & 4898307401048261248 & 04:35:11.52 & -22:45:32.8 &  2459512.76276 & 4x1800 & 16.72 & 17.22 & 16.06 & 23  & 83.4 $\pm$ 1.0 & 52  \\
Tur-3 & 5091015982252983680 & 04:10:37.44 & -20:28:59.3 & 2459498.79739 & 4x1800 & 16.71 & 16.91 & 16.31 & 29  & -11.9 $\pm$ 1.4 & 46 \\
WY-1 & 4725846184731540608 & 02:31:33.91 & -60:07:16.4  & 2459513.67965 & 2x1800 & 15.93 & 16.51 & 15.19 &  27  & 157.4 $\pm$ 1.1 & 56 \\
WY-2 & 4701367723002671744 & 02:25:47.86 & -62:09:45.1  & 2459498.67202 & 4x1800 & 16.70 & 17.16 & 16.06 &  14 & 174.6 $\pm$ 1.0 & 51 \\
WY-3 & 4725852300765007104 & 02:33:35.63 & -59:55:52.0  & 2459489.67301 & 4x1800  & 16.90 & 17.36 & 16.26 & 20  & 168.7 $\pm$ 1.4 & 53
\enddata
\end{deluxetable*}

\begin{deluxetable*}{ccrrrrrrrrrrrrc}

\tablecaption{\label{tab:uncertainties} Stellar Model Atmosphere Parameters}
\tablehead{& \colhead{} & \colhead{$T_{\rm eff}$} & \colhead{$\log g$} & \colhead{$\xi$} & \colhead{$\mathrm{[Fe/H]}$}\\
\colhead{} & \colhead{} & \colhead{(K)} &\colhead{(cgs)}  & \colhead{(km~s$^{-1}$)} & \colhead{(dex)}   }
\startdata
\multirow{3}{*}{Tur-1} 
& Value & 4924 & 1.84 & 2.21 & -2.42   \\
& Statistical Uncertainties & \nodata & 0.06 & 0.06 & 0.11   \\
& Systematic Uncertainties & 79 & 0.28 & 0.08 & 0.09 \\
\hline
\multirow{ 3}{*}{Tur-2}
& Value & 4963 & 2.20 & 2.08 & -2.37  \\
& Statistical Uncertainties & \nodata & 0.06 & 0.07 & 0.12 \\
& Systematic Uncertainties & 85 & .36 & 0.24 & 0.11  \\
\hline
\multirow{3}{*}{Tur-3}
& Value & 6756 & 1.83 & 3.50 & -2.53  \\
& Statistical Uncertainties & \nodata & 0.16 & 0.37 & 0.13   \\
& Systematic Uncertainties & 115 & 0.24 & 0.15 & 0.06  \\
\hline
\multirow{3}{*}{WY-1}
& Value & 4587 & 1.40 & 2.43 & -2.32  \\
& Statistical Uncertainties & \nodata & 0.05 & 0.06 & 0.10  \\
& Systematic Uncertainties & 70 & 0.30 & 0.14 & 0.07  \\
\hline
\multirow{3}{*}{WY-2}
& Value & 4847 & 1.36 & 2.28 & -2.38 \\
& Statistical Uncertainties & \nodata & 0.06 & 0.06 & 0.10 \\
& Systematic Uncertainties & 146 & 0.52 & 0.08 & 0.18 \\
\hline
\multirow{3}{*}{WY-3}
& Value & 4949 & 1.48 & 2.27 & -2.34 \\
& Statistical Uncertainties & \nodata & 0.05 & 0.06 & 0.12  \\
& Systematic Uncertainties & 70 & 0.30 & 0.03 & 0.08  \\

\enddata
\end{deluxetable*}

\begin{deluxetable*}{lccccc}
\caption{Tur-3 \label{tab:tur}}
\tablehead{Exposure & Phase & $V_{\rm helio}$ & $T_{\rm eff}$ &  Avg. $T_{\rm eff}$ & Avg. $V_{\rm helio}$  \\ & & (km~s$^{-1}$) & (K)  & (K) & (km~s$^{-1}$) }
\startdata
1 & 0.12 & -13.5 $\pm$ 2.0  & 6851 $\pm$ 200 & \multirow{3}{*}{6756 $\pm$ 115} & \multirow{3}{*}{-11.9 $\pm$ 1.4}\\
2 & 0.15 & -12.9 $\pm$ 2.4 & 6754 $\pm$ 200\\
3 & 0.17 & -9.3  $\pm$ 2.7  & 6663 $\pm$ 200 \\
\enddata
\end{deluxetable*}

\section{Stellar Parameter and Abundance Analysis \label{sec:stellar parameters and abundance analysis}}
 
Using the program Spectroscopy Made Hard(er) (\code{SMHR}\footnote{\url{https://github.com/andycasey/smhr}}; \citealt{casey2025}), abundances were derived from equivalent width (EW) measurements and spectral synthesis. \code{SMHR} runs the radiative transfer code \code{MOOG}\citep{sneden1973,sobeck2011} assuming local thermodynamical equilibrium (LTE). Line lists were adopted from \cite{ji2020a} and supplemented with line lists generated from linemake\footnote{\url{https://github.com/vmplacco/linemake}} \citep{placco2021} for neutron-capture elements. All line lists include hyperfine structure and isotopic shifts where applicable, and Solar abundances were taken from \citet{asplund2009}. One dimensional (1D) $\alpha$-enhanced ($\mathrm{[\alpha/Fe]}= +0.4$) ATLAS model atmospheres \citep{castelli2003} were used as input. For all neutron-capture elements, we used the $r$-process isotopic ratios from \citet{sneden2008}. Atomic data, EWs, and individual abundances for lines used in the analysis of Turranburra and Willka Yaku are listed in Tables \ref{tab:atomic_data_tur} and \ref{tab:atomic_data_wy}, respectively.

The effective temperatures ($T_{\rm eff}$) for Tur-1, Tur-2, and all three WY stars were determined photometrically using $Gaia$ and 2MASS $K$ magnitudes. The $Gaia$ $G$, $BP$, and $RP$ photometry for each star were obtained from $Gaia$ DR3 \citep{gaia2022}, with $K$ magnitudes taken from 2MASS \citep{cutri2003}. Reddening values, $E(B-V)$, were sourced from \cite{schlafly2011}. The $Gaia$ magnitudes were de-reddened following the process in \cite{gaia2018}, and de-reddened $K$ magnitudes were found using the extinction and reddening values from \cite{schlafly2011}. The de-reddened photometry was then used to determine the $T_{\mathrm{eff}}$ of each star following the color-$\mathrm{[Fe/H]}$-$T_{\mathrm{eff}}$ relations of \cite{mucciarelli2021}. The final de-reddened magnitudes and colors are listed in Table \ref{tab:temperatures} along with the derived $T_{\rm eff}$ from each color band. The final adopted temperature is the average across all color bands. Following the determination of $T_{\rm eff}$, $\log{g}$ was determined from the ionization balance between the \ion{Fe}{1} and \ion{Fe}{2} lines and $\xi$ were determined by removing any trend in line abundances with reduced EW for the \ion{Fe}{1}. EWs were measured by fitting Gaussian profiles to absorption features in the continuum-normalized spectra. Finally, the model metallicity is taken as the mean abundances of all measured lines for \ion{Fe}{1} and \ion{Fe}{2}. 

The uncertainty for each temperature derived from the different color relations, listed in Table \ref{tab:temperatures}, accounts for the uncertainty from the spread in Fe abundances, as well as the uncertainties provided by \cite{mucciarelli2021} for the given color band. This uncertainty was added in quadrature to the standard deviation ($\sigma_{std}$ in Table \ref{tab:temperatures}) of the temperatures across the color relations to determine the final uncertainty in $T_{\mathrm{eff}}$. To estimate the associated uncertainty on $\log{g}$, $\mathrm{[Fe/H]}$, and $\xi$, we offset the temperature by the systematic uncertainty in SMHR, re-derived these stellar parameters, and then took the difference as the systematic uncertainty. Statistical uncertainties on $\log{g}$, $\mathrm{[Fe/H]}$, and $\xi$ were determined by varying each parameter to match the standard deviation of \ion{Fe}{1} lines as listed in Table \ref{tab:uncertainties}. The final stellar parameters and their associated uncertainties are listed in Table \ref{tab:uncertainties}. 

Since Tur-3 is an RR Lyrae (RRL) star, the parameter determination was done following the procedure outlined by \cite{ji2020b} and \cite{for2011}, to account for the radial velocity variations. Three exposures for this star were used, and the phase for each exposure was determined. The phase was calculated using the epoch of maximum light and period from $Gaia$. The radial velocity for each exposure was then determined through an order-by-order cross-correlation with a spectrum of HD 122563, and care was taken to verify that no variations in velocity were present across the different wavelength regions. Each exposure was then shifted to rest-frame, and each order was co-added in SMHR. The $T_{\rm eff}$ was determined using the Temperature-Phase Relationships from \cite{for2011}. The $\log{g}$, $\mathrm{[Fe/H]}$, and $\xi$ were determined the same way as the other stars. EWs were measured using the line list from \cite{ji2020a} supplemented with the line list from \cite{for2010}. The phase, $V_{\rm helio}$, $T_{\mathrm{eff}}$, and the average $T_{\mathrm{eff}}$ and $V_{\rm helio}$ are listed in Table \ref{tab:tur}.

\section{Results \label{sec:results}}
\begin{figure*}
\centering
\includegraphics[width=\linewidth]{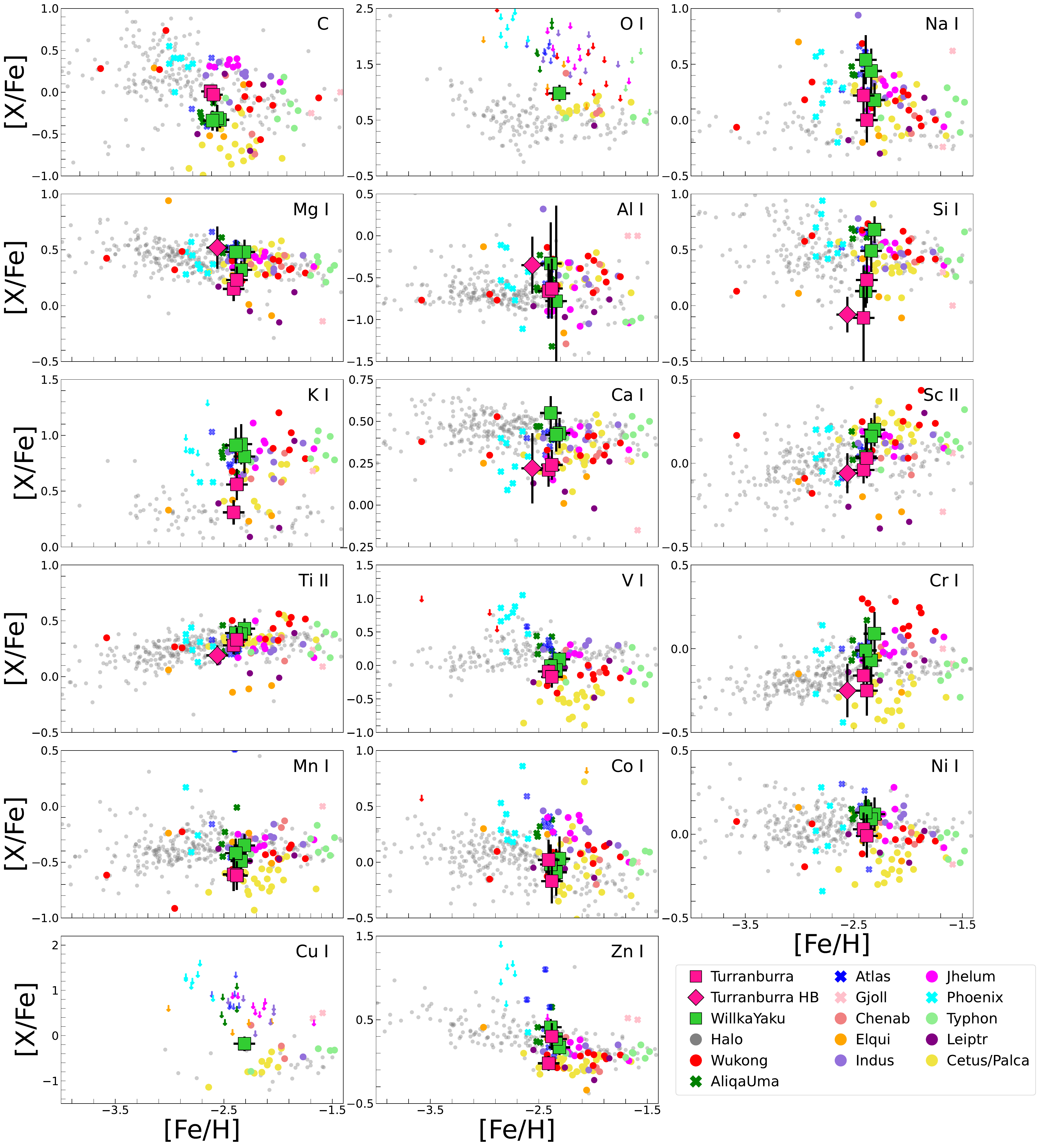}
\caption{\label{Fig:abundances} $\mathrm{[X/Fe]}$ derived abundances for the stars in Turranburra (Pink squares, RRL star is pink diamond) and Willka Yaku (green squares) compared to abundances of stars in other streams (colored circles represent streams with DG progenitors and colored X's represent streams with GC progenitors, see text for references), and stars from the MW halo (gray dots; \citealt{roederer2014}).}
\end{figure*}

For each of the three stars in Turranburra and Willka Yaku, abundances have been derived for a subset of elements spanning from C to Dy. The derived elemental abundances for elements with $Z\leq 30$ for the Turranburra and Willka Yaku stellar streams are shown in Figure \ref{Fig:abundances}, and are compared to other stream stars with high-resolution spectra from \SSSSS and MW halo stars \citep{roederer2014}. Data for the stellar streams come from the following sources: ATLAS, Aliqa Uma, Chenab, Elqui, Indus, Jhelum, and Phoenix\citep{ji2020a}; Typhon \citep{ji2023b}; Wukong \citep{limberg2024}; Indus \citep{hansen2021}; Leiptr \citep{atzberger2024}; and Palca/Cetus \citep{sitnova2024}. Tables \ref{tab:tur_abun} and \ref{tab:wy_abun} list the final $\log{\epsilon}$ abundances, the number of lines used (N) for each species, $\mathrm{[X/H]}$, $\mathrm{[X/Fe]}$, and associated uncertainties. Variations in the number of elements derived per star arise from differences in the quality of the spectra. Tables \ref{tab:tur_uncertainties} and \ref{tab:wy_uncertainties} list the abundance uncertainties arising from stellar parameter uncertainties, further detailed in \cite{ji2020a}. The final uncertainties were calculated using the method described by \cite{ji2020a}, which uses a signal-to-noise weighted mean to determine the final abundances. Both statistical and systematic uncertainties associated with stellar parameters were propagated to estimate the uncertainties for each elemental abundance.

\begin{figure*}
\centering
\includegraphics[width=\linewidth]{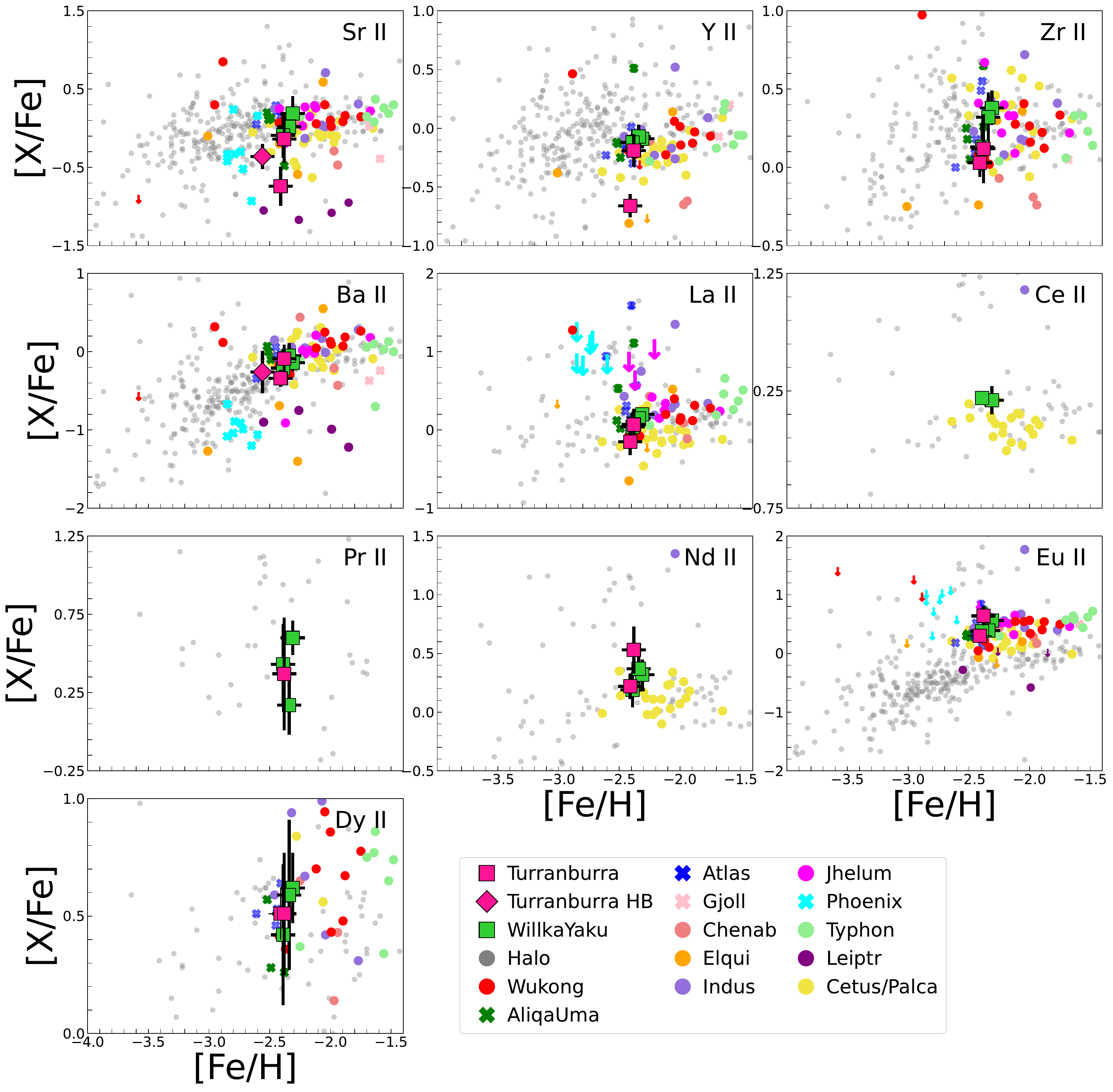}
\caption{\label{Fig:neutroncaptureelements} $\mathrm{[X/Fe]}$ derived neutron-capture element abundances for the Turranburra and Willka Yaku stars compared to abundances of other stellar streams and MW halo stars (see Fig.~\ref{Fig:abundances} for legend). }
\end{figure*}

\subsection{$\alpha$ elements}
Abundances for the $\alpha$-elements, O, Mg, Si, and Ca, were mostly derived from EWs, with the exception of Si, where spectral synthesis was used due to blended features. An O abundance could only be derived in one of the six stars, namely WY-1 ($\mathrm{[O I/Fe]= 0.98 \pm 0.15}$), while Mg, Si, and Ca abundances are derived for all six stars.  For Tur-1, we note that the abundances of the two Si lines employed at 4102~\angstrom and 3906~\angstrom to derive the Si abundance significantly disagree, resulting in the large uncertainty reported. As can be seen in Figure \ref{Fig:abundances}, the $\alpha$-element abundances we derive in Willka Yaku and Turranburra generally follow the $\alpha$-element abundance signature of other streams and stars in the MW halo.

\setlength{\tabcolsep}{3pt}
\begin{deluxetable*}{crrrrrr|rrrrrr|rrrrrc}
\tablecaption{\label{tab:tur_abun}Turranburra Abundances}
\tablehead{
\colhead{} & \multicolumn{6}{c}{Tur-1} &\multicolumn{6}{c}{Tur-2} &\multicolumn{6}{c}{Tur-3}  \\
\colhead{Species} & \colhead{N} &  \colhead{$\log\epsilon$(X)} & \colhead{$\mathrm{[X/H]}$} & \colhead{$\sigma_{\mathrm{[X/H]}}$} & \colhead{$\mathrm{[X/Fe]}$}& \colhead{$\sigma_{\mathrm{[X/Fe]}}$} &  \colhead{N} &  \colhead{$\log\epsilon$(X)} & \colhead{$\mathrm{[X/H]}$} & \colhead{$\sigma_{\mathrm{[X/H]}}$} & \colhead{$\mathrm{[X/Fe]}$} & \colhead{$\sigma_{\mathrm{[X/Fe]}}$} &  \colhead{N} &  \colhead{$\log\epsilon$(X)} & \colhead{$\mathrm{[X/H]}$} & \colhead{$\sigma_{\mathrm{[X/H]}}$} & \colhead{$\mathrm{[X/Fe]}$} & \colhead{$\sigma_{\mathrm{[X/Fe]}}$}  \\
\colhead{} & \colhead{} & \colhead{(dex)} & \colhead{(dex)} & \colhead{(dex)} &\colhead{(dex)} & \colhead{(dex)}  & \colhead{} & \colhead{(dex)} & \colhead{(dex)} & \colhead{(dex)} & \colhead{(dex)} & \colhead{(dex)}  & \colhead{} & \colhead{(dex)} & \colhead{(dex)} & \colhead{(dex)} & \colhead{(dex)} & \colhead{(dex)}}
\startdata
\ion{C (CH)}{0}	&	3	&	$+$6.03	&	$-$2.40	&	0.08	&	$+$0.01	&	0.08	&	3	&	$+$6.02	&	$-$2.41	&	0.13	&	$-$0.03	&	0.12	&	\nodata	&	\nodata	&	\nodata	&	\nodata	&	\nodata	&	\nodata	\\
\ion{CH$_{corr}$}{0}	&	\nodata	&	$+$6.28	&	$-$2.18	&	\nodata	&	$+$0.23	&	\nodata	&	\nodata	&	$+$6.07	&	$-$2.39 	&	\nodata	&	$-$0.01	&	\nodata	&	\nodata	&	\nodata	&	\nodata	&	\nodata	&	\nodata	&	\nodata	\\
\ion{Na}{1}	&	2	&	$+$4.04	&	$-$2.20	&	0.20	&	$+$0.22	&	0.20	&	2	&	$+$3.86	&	$-$2.38	&	0.20	&	0.00	&	0.20	&	\nodata	&	\nodata	&	\nodata	&	\nodata	&	\nodata	&	\nodata	\\
\ion{Mg}{1}	&	5	&	$+$5.33	&	$-$2.27	&	0.11	&	$+$0.15	&	0.11	&	5	&	$+$5.47	&	$-$2.15	&	0.12	&	$+$0.23	&	0.12	&	2	&	5.56	&	-2.04	&	0.25	&	0.52	&	0.19	\\
\ion{Al}{1}	&	2	&	$+$3.37	&	$-$3.08	&	0.33	&	$-$0.66	&	0.33	&	2	&	$+$3.44	&	$-$3.01	&	0.37	&	$-$0.63	&	0.36	&	1	&	3.54	&	-2.91	&	0.37	&	-0.35	&	0.34	\\
\ion{Si}{1}	&	2	&	$+$4.98	&	$-$2.53	&	0.47	&	$-$0.11	&	0.47	&	1	&	$+$5.36	&	$-$2.15	&	0.22	&	$+$0.23	&	0.21	&	1	&	4.88	&	-2.63	&	0.23	&	-0.08	&	0.16	\\
\ion{K}{1}	&	1	&	$+$2.93	&	$-$2.10	&	0.11	&	$+$0.31	&	0.11	&	2	&	$+$3.23	&	$-$1.82	&	0.14	&	$+$0.56	&	0.14	&	\nodata	&	\nodata	&	\nodata	&	\nodata	&	\nodata	&	\nodata	\\
\ion{Ca}{1}	&	15	&	$+$4.13	&	$-$2.21	&	0.10	&	$+$0.21	&	0.10	&	17	&	$+$4.22	&	$-$2.14	&	0.09	&	$+$0.24	&	0.09	&	1	&	4.00	&	-2.34	&	0.29	&	0.22	&	0.21	\\
\ion{Sc}{2}	&	8	&	$+$0.69	&	$-$2.46	&	0.10	&	$-$0.04	&	0.08	&	7	&	$+$0.80	&	$-$2.35	&	0.14	&	$+$0.03	&	0.12	&	2	&	0.53	&	-2.62	&	0.16	&	-0.06	&	0.12	\\
\ion{Ti}{1}	&	19	&	$+$2.64	&	$-$2.31	&	0.11	&	$+$0.11	&	0.11	&	15	&	$+$2.76	&	$-$2.23	&	0.14	&	$+$0.15	&	0.13	&	\nodata	&	\nodata	&	\nodata	&	\nodata	&	\nodata	&	\nodata	\\
\ion{Ti}{2}	&	21	&	$+$2.81	&	$-$2.14	&	0.09	&	$+$0.28	&	0.08	&	28	&	$+$2.98	&	$-$2.05	&	0.12	&	$+$0.33	&	0.11	&	14	&	2.58	&	-2.37	&	0.12	&	0.19	&	0.09	\\
\ion{V}{1}	&	2	&	$+$1.43	&	$-$2.50	&	0.11	&	$-$0.09	&	0.11	&	2	&	$+$1.39	&	$-$2.55	&	0.16	&	$-$0.17	&	0.16	&	\nodata	&	\nodata	&	\nodata	&	\nodata	&	\nodata	&	\nodata	\\
\ion{V}{2}	&	2	&	$+$1.67	&	$-$2.26	&	0.07	&	$+$0.15	&	0.08	&	2	&	$+$1.96	&	$-$1.97	&	0.15	&	$+$0.41	&	0.15	&	\nodata	&	\nodata	&	\nodata	&	\nodata	&	\nodata	&	\nodata	\\
\ion{Cr}{1}	&	12	&	$+$3.07	&	$-$2.57	&	0.14	&	$-$0.16	&	0.13	&	6	&	$+$3.05	&	$-$2.63	&	0.16	&	$-$0.25	&	0.15	&	2	&	2.83	&	-2.81	&	0.24	&	-0.25	&	0.16	\\
\ion{Cr}{2}	&	4	&	$+$3.26	&	$-$2.38	&	0.07	&	$+$0.03	&	0.07	&	3	&	$+$3.32	&	$-$2.38	&	0.10	&	0.00	&	0.09	&	1	&	3.27	&	-2.37	&	0.09	&	0.19	&	0.08	\\
\ion{Mn}{1}	&	7	&	$+$2.40	&	$-$3.03	&	0.15	&	$-$0.61	&	0.15	&	6	&	$+$2.43	&	$-$3.00	&	0.12	&	$-$0.62	&	0.13	&	\nodata	&	\nodata	&	\nodata	&	\nodata	&	\nodata	&	\nodata	\\
\ion{Fe}{1}	&	75	&	$+$5.09	&	$-$2.41	&	0.04	&	0.00	&	0.00	&	63	&	$+$5.12	&	$-$2.38	&	0.06	&	0.00	&	0.00	&	16	&	4.94	&	-2.56	&	0.18	&	0.00	&	0.00	\\
\ion{Fe}{2}	&	14	&	$+$5.08	&	$-$2.42	&	0.08	&	$-$0.01	&	0.00	&	14	&	$+$5.16	&	$-$2.35	&	0.10	&	$+$0.03	&	0.00	&	10	&	4.99	&	-2.51	&	0.12	&	0.05	&	0.00	\\
\ion{Co}{1}	&	6	&	$+$2.59	&	$-$2.40	&	0.18	&	$+$0.02	&	0.18	&	6	&	$+$2.44	&	$-$2.55	&	0.20	&	$-$0.17	&	0.20	&	\nodata	&	\nodata	&	\nodata	&	\nodata	&	\nodata	&	\nodata	\\
\ion{Ni}{1}	&	12	&	$+$3.84	&	$-$2.38	&	0.10	&	$+$0.03	&	0.10	&	8	&	$+$3.87	&	$-$2.39	&	0.13	&	$-$0.01	&	0.13	&	\nodata	&	\nodata	&	\nodata	&	\nodata	&	\nodata	&	\nodata	\\
\ion{Zn}{1}	&	2	&	$+$2.13	&	$-$2.43	&	0.09	&	$-$0.02	&	0.09	&	2	&	$+$2.52	&	$-$2.08	&	0.16	&	$+$0.30	&	0.16	&	\nodata	&	\nodata	&	\nodata	&	\nodata	&	\nodata	&	\nodata	\\
\ion{Sr}{2}	&	2	&	$-$0.29	&	$-$3.16	&	0.30	&	$-$0.74	&	0.25	&	2	&	$+$0.35	&	$-$2.52	&	0.35	&	$-$0.14	&	0.30	&	2	&	-0.05	&	-2.92	&	0.23	&	-0.36	&	0.16	\\
\ion{Y}{2}	&	2	&	$-$0.86	&	$-$3.07	&	0.12	&	$-$0.66	&	0.10	&	5	&	$-$0.36	&	$-$2.57	&	0.16	&	$-$0.19	&	0.14	&	\nodata	&	\nodata	&	\nodata	&	\nodata	&	\nodata	&	\nodata	\\
\ion{Zr}{2}	&	1	&	$+$0.20	&	$-$2.38	&	0.09	&	$+$0.03	&	0.09	&	1	&	$+$0.32	&	$-$2.26	&	0.21	&	$+$0.12	&	0.22	&	\nodata	&	\nodata	&	\nodata	&	\nodata	&	\nodata	&	\nodata	\\
\ion{Ba}{2}	&	5	&	$-$0.58	&	$-$2.76	&	0.14	&	$-$0.34	&	0.11	&	5	&	$-$0.29	&	$-$2.47	&	0.21	&	$-$0.09	&	0.18	&	1	&	-0.64	&	-2.82	&	0.30	&	-0.26	&	0.27	\\
\ion{La}{2}	&	2	&	$-$1.47	&	$-$2.57	&	0.19	&	$-$0.15	&	0.17	&	2	&	$-$1.21	&	$-$2.31	&	0.21	&	$+$0.07	&	0.20	&	\nodata	&	\nodata	&	\nodata	&	\nodata	&	\nodata	&	\nodata	\\
\ion{Pr}{2}	&	\nodata	&	\nodata	&	\nodata	&	\nodata	&	\nodata	&	\nodata	&	1	&	$-$1.29	&	$-$2.01	&	0.39	&	$+$0.37	&	0.36	&	\nodata	&	\nodata	&	\nodata	&	\nodata	&	\nodata	&	\nodata	\\
\ion{Nd}{2}	&	1	&	$-$0.78	&	$-$2.20	&	0.12	&	$+$0.22	&	0.11	&	2	&	$-$0.43	&	$-$1.85	&	0.20	&	$+$0.53	&	0.20	&	\nodata	&	\nodata	&	\nodata	&	\nodata	&	\nodata	&	\nodata	\\
\ion{Eu}{2}	&	2	&	$-$1.59	&	$-$2.11	&	0.11	&	$+$0.30	&	0.09	&	3	&	$-$1.22	&	$-$1.73	&	0.18	&	$+$0.64	&	0.16	&	\nodata	&	\nodata	&	\nodata	&	\nodata	&	\nodata	&	\nodata	\\
\ion{Dy}{2}	&	1	&	$-$0.80	&	$-$1.90	&	0.14	&	$+$0.51	&	0.11	&	1	&	$-$0.77	&	$-$1.87	&	0.28	&	$+$0.51	&	0.26	&	\nodata	&	\nodata	&	\nodata	&	\nodata	&	\nodata	&	\nodata	\\
\ion{Er}{2}	&	2	&	$-$1.13	&	$-$2.05	&	0.15	&	$+$0.36	&	0.15	&	2	&	$-$0.77	&	$-$1.69	&	0.25	&	$+$0.69	&	0.23	&	\nodata	&	\nodata	&	\nodata	&	\nodata	&	\nodata	&	\nodata	\\
\ion{Yb}{2}	&	1	&	$-$1.55	&	$-$2.39	&	0.23	&	$+$0.03	&	0.20	&	\nodata	&	\nodata	&	\nodata	&	\nodata	&	\nodata	&	\nodata	&	\nodata	&	\nodata	&	\nodata	&	\nodata	&	\nodata	&	\nodata	\\
\enddata
\end{deluxetable*}

\setlength{\tabcolsep}{3pt}
\begin{deluxetable*}{crrrrrr|rrrrrr|rrrrrc}

\tablecaption{ \label{tab:wy_abun} Willka Yaku Abundances}
\tablehead{
\colhead{} & \multicolumn{6}{c}{WY-1} & \multicolumn{6}{c}{WY-2} & \multicolumn{6}{c}{WY-3}  \\
\colhead{Species} & \colhead{N} & \colhead{$\log\epsilon$(X)} & \colhead{$\mathrm{[X/H]}$} & \colhead{$\sigma_{\mathrm{[X/H]}}$} & \colhead{$\mathrm{[X/Fe]}$} & \colhead{$\sigma_{\mathrm{[X/Fe]}}$} & \colhead{N} & \colhead{$\log\epsilon$(X)} & \colhead{$\mathrm{[X/H]}$} & \colhead{$\sigma_{\mathrm{[X/H]}}$} & \colhead{$\mathrm{[X/Fe]}$} & \colhead{$\sigma_{\mathrm{[X/Fe]}}$} & \colhead{N} & \colhead{$\log\epsilon$(X)} & \colhead{$\mathrm{[X/H]}$} & \colhead{$\sigma_{\mathrm{[X/H]}}$} & \colhead{$\mathrm{[X/Fe]}$} & \colhead{$\sigma_{\mathrm{[X/Fe]}}$} \\ 
\colhead{} & \colhead{} & \colhead{(dex)} & \colhead{(dex)} & \colhead{(dex)} &\colhead{(dex)} & \colhead{(dex)} & \colhead{} & \colhead{(dex)} & \colhead{(dex)} & \colhead{(dex)} & \colhead{(dex)} & \colhead{(dex)} & \colhead{} & \colhead{(dex)} & \colhead{(dex)} & \colhead{(dex)} & \colhead{(dex)} & \colhead{(dex)}
}
\startdata
\ion{C (CH)} {0}	&	3	&	$+$5.79	&	$-$2.64	&	0.09	&	$-$0.33	&	0.10	&	2	&	$+$5.70	&	$-$2.73	&	0.17	&	$-$0.34	&	0.16	&	2	&	$+$5.79	&	$-$2.64	&	0.12	&	$-$0.31	&	0.12	\\
\ion{CH$_{corr}$}{0}	 & 	\nodata 	& 	$+$6.39 	&	 $-$2.07	 & 	\nodata	 & 	$+$0.24	& 	\nodata	 &  	\nodata 	& 	$+$6.29	& 	$-$2.17	 &	 \nodata	 &	$+$0.22	&	 \nodata	 &  	\nodata 	& 	$+$6.31 &	 $-$2.15	&	 \nodata 	& 	$+$0.19	& 	\nodata  	\\
\ion{N (CN)}{0}	&	1	&	$+$6.01	&	$-$1.82	&	0.36	&	$+$0.49	&	0.36	&	\nodata	&	\nodata	&	\nodata	&	\nodata	&	\nodata	&	\nodata	&	\nodata	&	\nodata	&	\nodata	&	\nodata	&	\nodata	&	\nodata	\\
\ion{O}{1}	&	1	&	$+$7.36	&	$-$1.33	&	0.14	&	$+$0.98	&	0.15	&	\nodata	&	\nodata	&	\nodata	&	\nodata	&	\nodata	&	\nodata	&	\nodata	&	\nodata	&	\nodata	&	\nodata	&	\nodata	&	\nodata	\\
\ion{Na}{1}	&	3	&	$+$4.11	&	$-$2.13	&	0.15	&	$+$0.18	&	0.15	&	2	&	$+$4.39	&	$-$1.85	&	0.20	&	$+$0.54	&	0.20	&	2	&	$+$4.35	&	$-$1.89	&	0.22	&	$+$0.44	&	0.22	\\
\ion{Mg}{1}	&	4	&	$+$5.77	&	$-$1.83	&	0.11	&	$+$0.48	&	0.11	&	9	&	$+$5.69	&	$-$1.91	&	0.07	&	$+$0.48	&	0.08	&	6	&	$+$5.59	&	$-$2.01	&	0.11	&	$+$0.32	&	0.11	\\
\ion{Al}{1}	&	\nodata	&	\nodata	&	\nodata	&	\nodata	&	\nodata	&	\nodata	&	1	&	$+$3.73	&	$-$2.72	&	1.15	&	$-$0.33	&	1.14	&	1	&	$+$3.34	&	$-$3.11	&	0.50	&	$-$0.78	&	0.49	\\
\ion{Si}{1}	&	4	&	$+$5.88	&	$-$1.63	&	0.12	&	$+$0.68	&	0.12	&	1	&	$+$5.25	&	$-$2.26	&	0.19	&	$+$0.13	&	0.19	&	1	&	$+$5.67	&	$-$1.84	&	0.16	&	$+$0.49	&	0.15	\\
\ion{K}{1}	&	2	&	$+$3.53	&	$-$1.50	&	0.15	&	$+$0.81	&	0.15	&	2	&	$+$3.55	&	$-$1.48	&	0.19	&	$+$0.91	&	0.18	&	2	&	$+$3.62	&	$-$1.41	&	0.16	&	$+$0.92	&	0.16	\\
\ion{Ca}{1}	&	20	&	$+$4.46	&	$-$1.88	&	0.09	&	$+$0.43	&	0.09	&	15	&	$+$4.49	&	$-$1.85	&	0.10	&	$+$0.55	&	0.10	&	12	&	$+$4.42	&	$-$1.92	&	0.10	&	$+$0.42	&	0.10	\\
\ion{Sc}{2}	&	6	&	$+$1.04	&	$-$2.11	&	0.11	&	$+$0.20	&	0.10	&	6	&	$+$0.80	&	$-$2.35	&	0.14	&	$+$0.04	&	0.11	&	7	&	$+$0.98	&	$-$2.17	&	0.11	&	$+$0.16	&	0.11	\\
\ion{Ti}{1}	&	28	&	$+$2.96	&	$-$1.99	&	0.08	&	$+$0.32	&	0.09	&	14	&	$+$2.87	&	$-$2.08	&	0.12	&	$+$0.31	&	0.12	&	15	&	$+$2.97	&	$-$1.98	&	0.13	&	$+$0.35	&	0.13	\\
\ion{Ti}{2}	&	33	&	$+$3.07	&	$-$1.99	&	0.09	&	$+$0.43	&	0.09	&	25	&	$+$2.95	&	$-$2.00	&	0.09	&	$+$0.39	&	0.10	&	32	&	$+$3.00	&	$-$1.95	&	0.07	&	$+$0.38	&	0.07	\\
\ion{V}{1}	&	3	&	$+$1.71	&	$-$2.22	&	0.13	&	$+$0.09	&	0.12	&	2	&	$+$1.53	&	$-$2.40	&	0.14	&	$-$0.01	&	0.14	&	2	&	$+$1.53	&	$-$2.40	&	0.17	&	$-$0.07	&	0.17	\\
\ion{V}{2}	&	2	&	$+$1.92	&	$-$2.01	&	0.11	&	$+$0.30	&	0.10	&	2	&	$+$1.59	&	$-$2.34	&	0.15	&	$+$0.06	&	0.13	&	2	&	$+$1.63	&	$-$2.30	&	0.10	&	$+$0.03	&	0.10	\\
\ion{Cr}{1}	&	10	&	$+$3.21	&	$-$2.43	&	0.15	&	$-$0.12	&	0.15	&	5	&	$+$3.26	&	$-$2.38	&	0.15	&	$+$0.02	&	0.14	&	8	&	$+$3.12	&	$-$2.52	&	0.13	&	$-$0.19	&	0.13	\\
\ion{Cr}{2}	&	4	&	$+$3.47	&	$-$2.17	&	0.08	&	$+$0.14	&	0.08	&	3	&	$+$3.33	&	$-$2.31	&	0.09	&	$+$0.09	&	0.07	&	3	&	$+$3.17	&	$-$2.47	&	0.07	&	$-$0.14	&	0.07	\\
\ion{Mn}{1}	&	7	&	$+$2.77	&	$-$2.66	&	0.10	&	$-$0.35	&	0.10	&	5	&	$+$2.62	&	$-$2.81	&	0.12	&	$-$0.42	&	0.12	&	4	&	$+$2.61	&	$-$2.82	&	0.13	&	$-$0.49	&	0.13	\\
\ion{Fe}{1}	&	71	&	$+$5.19	&	$-$2.31	&	0.04	&	0.00	&	0.00	&	55	&	$+$5.11	&	$-$2.39	&	0.05	&	0.00	&	0.00	&	66	&	$+$5.16	&	$-$2.34	&	0.05	&	0.00	&	0.00	\\
\ion{Fe}{2}	&	14	&	$+$5.18	&	$-$2.32	&	0.08	&	$-$0.01	&	0.00	&	10	&	$+$5.12	&	$-$2.38	&	0.11	&	$+$0.02	&	0.00	&	19	&	$+$5.16	&	$-$2.34	&	0.06	&	0.00	&	0.00	\\
\ion{Co}{1}	&	3	&	$+$2.71	&	$-$2.28	&	0.20	&	$+$0.03	&	0.20	&	2	&	$+$2.58	&	$-$2.41	&	0.21	&	$-$0.02	&	0.21	&	2	&	$+$2.56	&	$-$2.43	&	0.19	&	$-$0.09	&	0.18	\\
\ion{Ni}{1}	&	18	&	$+$4.03	&	$-$2.19	&	0.10	&	$+$0.12	&	0.10	&	6	&	$+$3.95	&	$-$2.27	&	0.10	&	$+$0.13	&	0.10	&	13	&	$+$3.97	&	$-$2.25	&	0.10	&	$+$0.09	&	0.10	\\
\ion{Cu}{1}	&	1	&	$+$1.70	&	$-$2.50	&	0.16	&	$-$ 0.18	&	0.16	&	\nodata	&	\nodata	&	\nodata	&	\nodata	&	\nodata	&	\nodata	&	\nodata	&	\nodata	&	\nodata	&	\nodata	&	\nodata	&	\nodata	\\
\ion{Zn}{1}	&	2	&	$+$2.42	&	$-$2.14	&	0.08	&	$+$0.17	&	0.09	&	2	&	$+$2.57	&	$-$1.99	&	0.12	&	$+$0.41	&	0.12	&	2	&	$+$2.50	&	$-$2.06	&	0.10	&	$+$0.27	&	0.10	\\
\ion{Sr}{2}	&	2	&	$+$0.75	&	$-$2.12	&	0.24	&	$+$0.19	&	0.22	&	2	&	$+$0.38	&	$-$2.49	&	0.29	&	$-$0.09	&	0.23	&	2	&	$+$0.56	&	$-$2.31	&	0.31	&	$+$0.02	&	0.29	\\
\ion{Y}{2}	&	3	&	$-$0.19	&	$-$2.40	&	0.10	&	$-$0.09	&	0.08	&	3	&	$-$0.30	&	$-$2.51	&	0.11	&	$-$0.12	&	0.10	&	3	&	$-$0.20	&	$-$2.41	&	0.12	&	$-$0.07	&	0.11	\\
\ion{Zr}{2}	&	1	&	$+$0.65	&	$-$1.93	&	0.10	&	$+$0.38	&	0.11	&	1	&	$+$0.32	&	$-$2.26	&	0.17	&	$+$0.13	&	0.16	&	1	&	$+$0.56	&	$-$2.01	&	0.16	&	$+$0.32	&	0.15	\\
\ion{Ba}{2}	&	5	&	$-$0.27	&	$-$2.45	&	0.19	&	$-$0.14	&	0.16	&	5	&	$-$0.42	&	$-$2.60	&	0.22	&	$-$0.21	&	0.17	&	5	&	$-$0.21	&	$-$2.39	&	0.19	&	$-$0.05	&	0.17	\\
\ion{La}{2}	&	4	&	$-$1.01	&	$-$2.11	&	0.08	&	$+$0.20	&	0.08	&	3	&	$-$1.25	&	$-$2.35	&	0.15	&	$+$0.04	&	0.13	&	3	&	$-$1.09	&	$-$2.19	&	0.11	&	$+$0.15	&	0.10	\\
\ion{Ce}{2}	&	2	&	$-$0.56	&	$-$2.14	&	0.12	&	$+$0.17	&	0.10	&	1	&	$-$0.62	&	$-$2.20	&	0.21	&	$+$0.19	&	0.17	&	\nodata	&	\nodata	&	\nodata	&	\nodata	&	\nodata	&	\nodata	\\
\ion{Pr}{2}	&	1	&	$-$0.99	&	$-$1.71	&	0.14	&	$+$0.60	&	0.11	&	1	&	$-$1.24	&	$-$1.96	&	0.24	&	$+$0.43	&	0.19	&	1	&	$-$1.45	&	$-$2.17	&	0.28	&	$+$0.17	&	0.26	\\
\ion{Nd}{2}	&	3	&	$-$0.57	&	$-$1.99	&	0.15	&	$+$0.32	&	0.13	&	2	&	$-$0.78	&	$-$2.20	&	0.23	&	$+$0.19	&	0.20	&	1	&	$-$0.54	&	$-$1.96	&	0.16	&	$+$0.37	&	0.15	\\
\ion{Sm}{2}	&	1	&	$-$0.83	&	$-$1.79	&	0.16	&	$+$0.52	&	0.15	&	1	&	$-$0.95	&	$-$1.91	&	0.21	&	$+$0.48	&	0.19	&	\nodata	&	\nodata	&	\nodata	&	\nodata	&	\nodata	&	\nodata	\\
\ion{Eu}{2}	&	2	&	$-$1.23	&	$-$1.75	&	0.10	&	$+$0.56	&	0.08	&	2	&	$-$1.49	&	$-$2.01	&	0.13	&	$+$0.38	&	0.09	&	2	&	$-$1.43	&	$-$1.96	&	0.11	&	$+$0.39	&	0.10	\\
\ion{Gd}{2}	&	1	&	$-$0.56	&	$-$1.63	&	0.19	&	$+$0.68	&	0.18	&	1	&	$-$0.70	&	$-$1.77	&	0.27	&	$+$0.62	&	0.29	&	\nodata	&	\nodata	&	\nodata	&	\nodata	&	\nodata	&	\nodata	\\
\ion{Dy}{2}     	&	2	&	$-$0.59	&	$-$1.69	&	0.17	&	$+$0.62	&	0.15	&	1	&	$-$0.87	&	$-$1.97	&	0.33	&	$+$0.42	&	0.32	&	2	&	$-$0.67	&	$-$1.77	&	0.31	&	$+$0.59	&	0.30	\\
\ion{Ho}{2}	&	2	&	$-$1.26	&	$-$1.74	&	0.20	&	$+$0.57	&	0.19	&	\nodata	&	\nodata	&	\nodata	&	\nodata	&	\nodata	&	\nodata	&	2	&	$-$1.30	&	$-$1.78	&	0.27	&	$+$0.56	&	0.25	\\
\ion{Er}{2}    	&	2	&	$-$1.07	&	$-$1.99	&	0.15	&	$+$0.32	&	0.15	&	\nodata	&	\nodata	&	\nodata	&	\nodata	&	\nodata	&	\nodata	&	2	&	$-$0.93	&	$-$1.85	&	0.18	&	$+$0.48	&	0.17	\\
\enddata
\end{deluxetable*}

\subsection{Carbon and odd Z elements}
Abundances for C, Al, and Sc were derived using spectral synthesis, while abundances for Na and K were derived using EWs. As can be seen in Figure \ref{Fig:abundances}, the abundances we derived for these elements in Willka Yaku and Turranburra again generally follow the signature seen in other streams and the MW halo. It should be noted that \cite{roederer2014} applied non-LTE corrections to the K and Na abundances for the halo sample used in Figure \ref{Fig:abundances}, creating a slight ($\sim$0.4dex) offset between the stream stars and the halo stars for K. The C abundances were derived from the CH bands around the regions of 4310~\angstrom, 4323~\angstrom, and 4370~\angstrom. For stars where no O abundance could be derived, we assumed a standard oxygen enhancement for metal-poor stars of $\mathrm{[OI/Fe]}$= 0.4 when deriving the C abundances. A N abundance from the CN band at 3880~\angstrom was derived for WY-1. To account for the evolutionary state of the star, the tool from \citet{placco2014} was used to obtain carbon corrections. Both the uncorrected (C(CH)) and corrected (CH$_{corr}$) C abundances can be found in Tables \ref{tab:tur_abun} and \ref{tab:wy_abun}. None of the stars in either Turranburra or Willka Yaku can be classified as carbon-enhanced metal-poor stars (CEMP; $\mathrm{[C/Fe]} >0.7$, \citealt{beers2005, aoki2007}). 


\subsection{Iron peak elements} 
Abundances for the iron-peak elements Ti, Cr, Ni, and Zn were determined using EWs, while those for V, Mn, Co, and Cu were derived through spectral synthesis to account for hyperfine structure. As shown in Figure \ref{Fig:abundances}, the resulting abundances for these elements in both Willka Yaku and Turranburra generally follow the iron-peak abundance trends observed in other stellar streams and MW halo stars. Additionally, our analysis reveals good agreement between neutral and ionized species of Ti, Cr, and V across all stars in both systems (see Tables \ref{tab:tur_abun} and \ref{tab:wy_abun}).

\subsection{Neutron-capture elements \label{sec:neutron capture elements}} 
\begin{figure}
\centering
\includegraphics[scale=0.5]{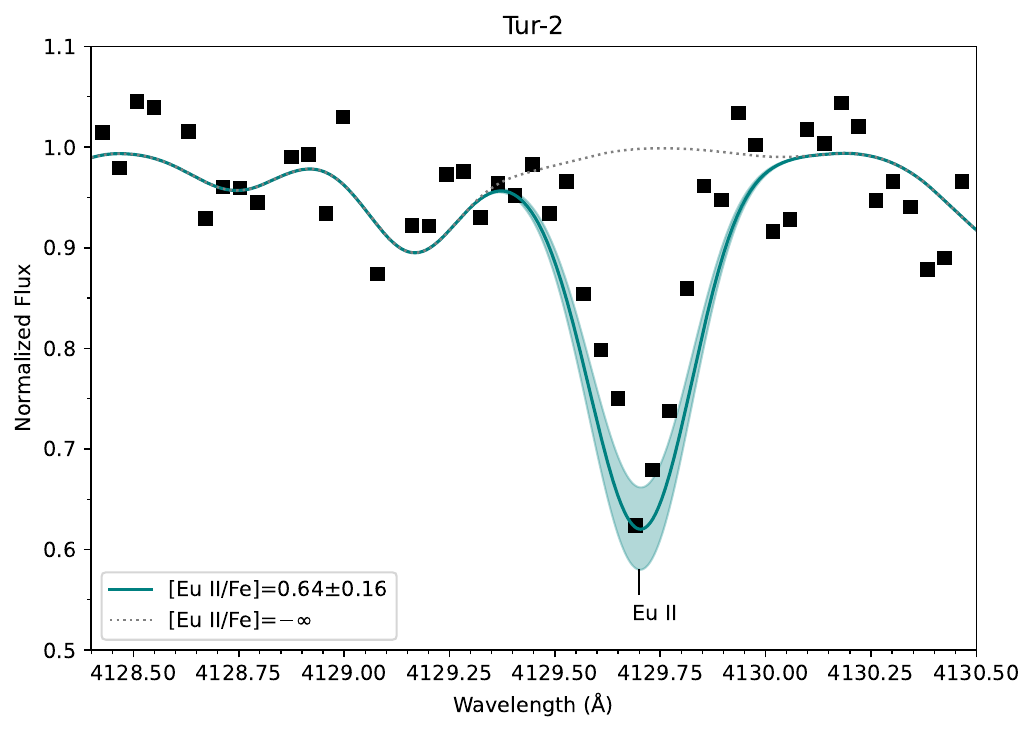}\\
\includegraphics[scale=0.5]{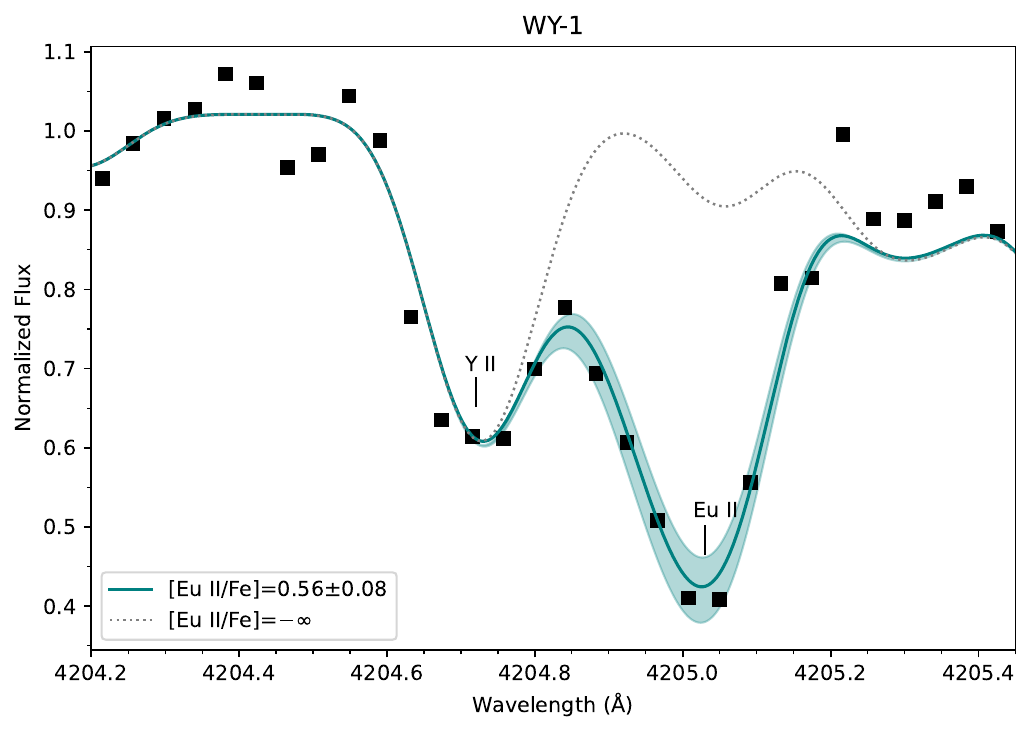}
\caption{Spectral synthesis of the 4129 \angstrom \ion{Eu}{2} for Tur-2 (top) and 4205 \angstrom WY-1 (bottom). Black squares are the observed spectrum, the dotted line shows a synthesis not including Eu, and the blue line shows a synthesis with the $\mathrm{[Eu II/Fe]}$ value derived for Tur-2 and WY-1. \label{Fig:eu_4205_synthesis} } 
\end{figure}

Abundances for the neutron-capture elements Sr, Y, Zr, Ba, La, Ce, Pr, Nd, Eu, and Dy were derived from spectral synthesis. The derived abundances for the neutron-capture elements in the two streams are shown in Figure \ref{Fig:neutroncaptureelements} and generally overlap with neutron-capture element abundances derived for other stream stars and MW halo stars. We note that only a few other stream stars have abundances derived for the heavy elements Ce, Pr, and Nd. Hence, this work significantly increases the available neutron-capture element abundance data for stream stars. Both Turranburra and Willka Yaku, along with stars in other streams, show a slight enhancement in Eu with Tur-1 and Tur-2 having $\mathrm{[Eu/Fe]}$ values of $0.30 \pm 0.09$ and $0.64 \pm 0.16$ respectively and WY-1, WY-2, and WY-3 having $\mathrm{[Eu/Fe]}$ values of $0.56 \pm 0.08$, $0.38 \pm 0.09$, and $0.39 \pm 0.10$ respectively. Figure \ref{Fig:eu_4205_synthesis} shows the synthesis of the Eu 4129~\angstrom and Eu 4205~\angstrom line for Tur-2 and WY-1 respectively. The $\mathrm{[Ba/Eu]}$ ratios of the stars are $-$0.70$\pm{0.24}$, $-0.59\pm{0.26}$, and $-0.44\pm{0.27}$ for WY-1, 2, and 3, respectively, and $-0.64\pm{0.20}$ and $-0.73 \pm{0.34}$ for Tur-1 and 2, consistent with enrichment by an $r$-process event \citep{sneden2008}.

\section{Discussion \label{sec:discussion}}

\subsection{Nature of stream progenitors}
An important step when using stellar streams to investigate the dynamical and chemical build-up of the MW is to determine the nature of the stream progenitors. Morphologically, stellar streams originating from DGs and GCs are different \citep{johnston2008, amorisco2015, erkal2016}. The GC streams are spatially thinner and dynamically cooler due to being less massive and having low-velocity dispersions of $\lesssim$ $5-10$ km~s$^{-1}$ compared to those formed from DGs \citep{helmi2020, tavangar2022,freeman2017}. However, in GCs, both the velocity dispersions and widths can increase for dynamical reasons \citep{malhan2021, li2022}. DG streams also generally exhibit a larger spread in metallicities ($\sigma_{[Fe/H]} \geq 0.20$; \citealt{li2022}) compared to GCs, due to more extended star formation. Further chemical signatures of the systems can also provide clues to the nature of the progenitor. For example, it is known that GCs have a spread in the light element abundances \citep{carretta2019} and there are anti-correlations such as Na-O and Mg-Al seen across many GCs \citep{carretta2009,carretta2010}. A chemical signature that can be seen in the classical DGs is that the more metal-rich stars exhibit lower $\mathrm{[\alpha/Fe]}$ ratios due to having a prolonged star formation history that allowed the chemical feedback from Type Ia supernovae to be incorporated into new generations of stars \citep{tolstoy2009}. This is a signature that has been seen in a few streams, such as the Wukong stellar stream \citep{limberg2024}, the Orphan-Chenab stream \citep{hawkins2023}, and the Helmi stream \citep{limberg2021}.


\subsubsection{Turranburra}
 The morphology of Turranburra described by \citet{shipp2018, shipp2019} suggests that the progenitor is a DG, whose mass is estimated as $\sim 1.8\times10^6$ M$_\odot$ from the stream width (288~pc) and the relations of \cite{erkal2016}. Additionally, \citet{li2022} determined a velocity dispersion for Turranburra of $\sigma_{vel}$= 19.7 km~s$^{-1}$.  It is noted in \citet{li2022} section 4.5, that this high dispersion could be due to processes that have heated the stream. A metallicity dispersion of $\sigma_{[Fe/H]}=0.39^{+0.12}_{-0.09}$ with $95\%$ confidence was also found by \citet{li2022} and is consistent with what is seen in classical DGs.
\cite{li2022} derived metallicities from the calcium triplet (CaT) for 22 stars in Turranburra and found an average CaT metallicity of $\overline{\mathrm{[Fe/H]}}_{CaT} = -2.18^{+0.13}_{-0.14}$. Specifically for the stars analyzed in this paper, \cite{li2022} found $\mathrm{[Fe/H]}= -2.58 \pm{0.13}$ and $-2.48 \pm{0.04}$ for Tur-1 and Tur-2, respectively; for comparison we find $\mathrm{[Fe/H]}=-2.41 \pm {0.04}$ and $-2.38 \pm{0.06}$ for Tur-1 and Tur-2 and a metallicity of $-2.56 \pm{0.18}$ for Tur-3, which was not in \cite{li2022}. From the analysis of the three stars presented in this paper, we find an average metallicity of $\mathrm{[Fe/H]= -2.45}$ and a standard deviation of $\sigma_{[Fe/H]}=0.07$. This value being lower than what was found by \cite{li2022} is likely due to the small sample size of three stars. This low average metallicity supports the idea that this stream comes from a low-mass DG. Another stellar stream that comes from a low-mass DG is Elqui, which is predicted to have a progenitor mass of $\sim 10^6 M_\odot$, and generally follows an abundance pattern similar to what is seen in Draco \citep{ji2020a}. Both Elqui and Draco have lower Mg abundances at higher metallicities, and Elqui has Solar level abundances for the neutron capture elements. Elqui and Turranburra show chemical similarities, but with only three stars in Turranburra, it is difficult to make a comprehensive comparison. However, as more chemical studies of streams are carried out, it appears that DG streams like Elqui and potentially Turranburra are beginning to fill in the mass range between UFD galaxies and the classical DGs.

\subsubsection{Willka Yaku}
The initial data collected for Willka Yaku showed it is a narrow stellar stream with a width of 127~pc and a suggested progenitor stellar mass of $\sim 1.4\times10^5$ M$_\odot$ \citep{shipp2018}. Later, \cite{li2022} found a low spread in velocity of $\sigma_{vel}$= 0.4 km~s$^{-1}$ and low metallicity dispersion of $\sigma_{[Fe/H]}=0.04^{+0.07}_{-0.02}$ with $95\%$ confidence for the Willka Yaku stars, implying that this stream's progenitor was a GC. \cite{li2022} derived an average CaT metallicity of $\overline{\mathrm{[Fe/H]}}_{CaT}= -2.1^{+0.8}_{-0.4}$ from nine stars. For the stars overlapping with this paper they found $\mathrm{[Fe/H]}= -2.02 \pm{0.04}, -2.44 \pm{0.05},$ and $-2.35 \pm{0.05}$ for WY-1, WY-2, and WY-3 respectively. With the high-resolution data for the three stars presented in this paper, we found $\mathrm{[Fe/H]}= -2.31 \pm{0.04}, -2.39 \pm{0.05}, $ and $-2.34 \pm{0.05}$ for WY-1, WY-2, and WY-3, respectively. We find an average metallicity of $\mathrm{[Fe/H]}= -2.35$ and a low standard deviation of the metallicities derived for the three stars of $\sim0.02$ dex in Willka Yaku, which agrees with the suggestion that this stream has a GC progenitor. As mentioned above, a trend commonly found in intact GCs is a spread in the lighter elements such as Na, Mg, and Al, due to anti-correlations of specific abundance pairs such as Na-O and Mg-Al. Multiple stellar populations, characterized by distinct chemical abundance patterns, are seen in almost all GCs \citep{milone2017, gratton2019} and have been identified in stellar streams before \citep{usman2024}. The fraction of second-population stars in a GC correlates with the mass of the system, with more massive GCs containing a higher fraction \citep{milone2017}. GCs with initial masses of $\sim10^{5.5}$ M$_\odot$ will have a second-population star fraction of $\sim50\%$ \citep{gratton2019, usman2024}, so given Willka Yaku's predicted progenitor mass of $\sim 1.4\times10^5$ M$_\odot$, the presence of second-population stars would be expected. Both the Na ($\mathrm{[Na/Fe]}= 0.18 \pm{0.15}, 0.54 \pm{0.20}$, and $0.44 \pm{0.22}$ for WY-1, 2, and 3 respectively) and Al ( $\mathrm{[Al/Fe]}= -0.33 \pm{1.14}$ and $-0.78 \pm{0.49}$ for WY-2, and 3 respectively) abundances are consistent within their uncertainties, showing that there is no measurable light element abundance spread between the three stars in Willka Yaku.

This could, on the other hand, point to the system being an open cluster, which are systems known to be chemically homogeneous ($\Delta\mathrm{[Fe/H]\lesssim 0.05}$ dex, \citealt{bland-hawthorn2010, krumholz2019}). However, with a sample of three stars, it is difficult to make conclusions on the presence of either signatures.

Willka Yaku is close in phase space to multiple systems, including the Palca/Cetus stream and NGC 5824 \citep{li2022}. \cite{bonaca2021} stated that it is likely that Willka Yaku was brought into the MW by the progenitor of the Cetus stream and \cite{li2022} found that it is possible the progenitors for both Willka Yaku and NGC 5824 fell into the MW together. It can be seen in Figures \ref{Fig:abundances} and \ref{Fig:neutroncaptureelements} that the Palca/Cetus stream (yellow dots) and Willka Yaku share some chemical similarities. However a larger sample is needed to confidently chemically link the systems.  

\subsection{$r$-process enhancement in stellar streams \label{sec:rprocess}}
To date, an $\mathrm{[Eu/Fe]}$ enhancement has been detected in almost all stellar streams, both originating from GCs and DGs. Eu is almost entirely produced in the $r$-process, the astrophysical production site(s) of which is still heavily debated \citep{cowan2021}. Investigating these abundance signatures in stellar streams can give us clues to the chemical enrichment history of these systems and the nucleosynthesis channels operating in the early universe, including the $r$-process. 

Many intact GCs show an enhancement in $r$-process elements, and some of those show a dispersion in the heavy elements (e.g., \citealt{zevin2019, roederer2011, kirby2023, sneden1997}); however, the source for this signature is not known. A potential explanation for this was discussed by \cite{kirby2023}, who investigated the $r$-process element signature of the GC M92 and found that a dispersion in $r$-process element abundances in this GC of $\sigma \sim 0.15$ indicates a first-generation population, while a lower dispersion suggests a second-generation population. Willka Yaku shows a low dispersion of $\sigma \sim 0.08$ in Eu, which, according to the threshold from \citet{kirby2023}, suggests the stars analyzed in this paper are part of a second-generation population. However, the low dispersion we determine is likely influenced by the small sample size (three stars). Further investigation of the spread of Eu abundance in the Willka Yaku system, as well as other GC streams, is thus needed to make a firm conclusion.

\begin{figure}
\centering
\includegraphics[width=\linewidth]{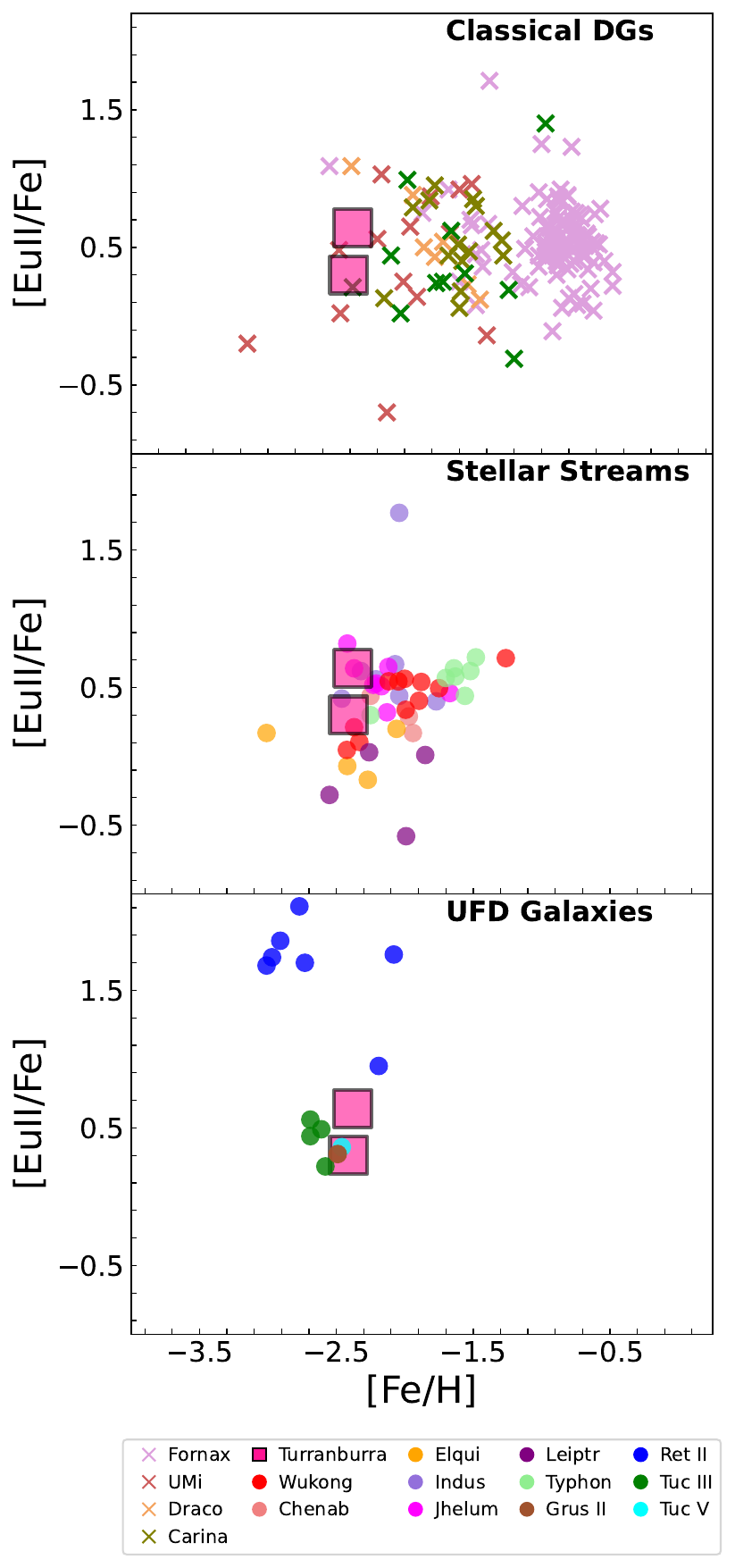}
\caption{\label{Fig:Eu_Fe} $\mathrm{[Eu II/Fe]}$ abundances for stars in classical DGs (\citealt{letarte2010, aoki2007_umi, cohen2010, shetrone2003, norris2017, venn2012, cohen2009}) (top), Turranburra and streams with DG progenitors (middle), and UFD galaxies (\citealt{ ji2016c, marshall2019, hansen2020, hansen2024}) (bottom) }
\end{figure}

\begin{figure*}
\centering
\includegraphics[width=0.85\textwidth]{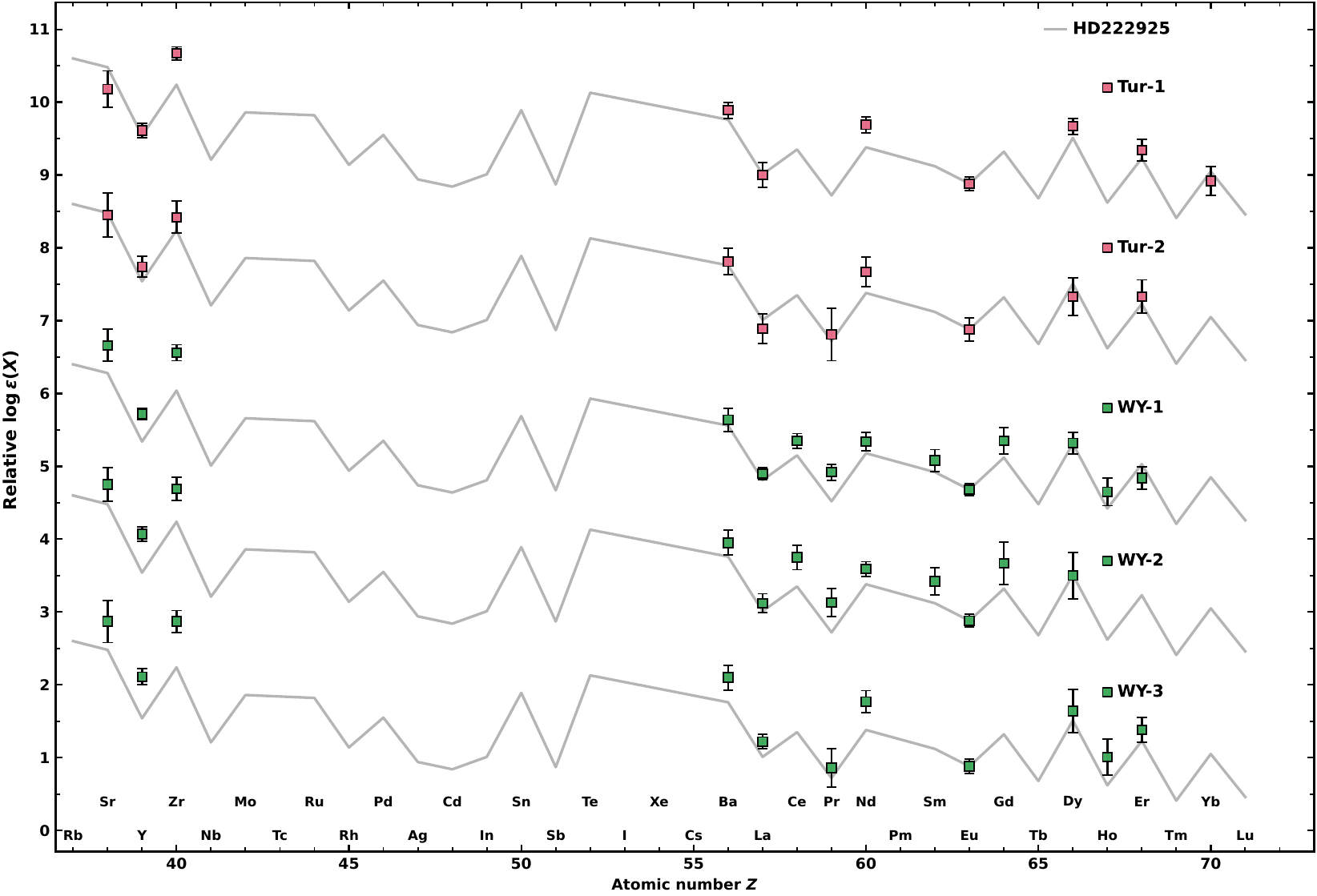}
\caption{\label{Fig:r_process_pattern}Absolute abundances for neutron-capture elements derived for Tur-1, Tur-2, WY-1, WY-2, and WY-3 (pink and green squares respectively) compared to the abundances of HD~222925 from \cite{roederer2022} (grey line). Each set of abundances for the stream star is scaled to Eu, and a further arbitrary offset is applied for visual clarity.}
\end{figure*}

 Many of the intact classical DGs show a mild enhancement in $\mathrm{[Eu/Fe]}$. For example, the Ursa Minor, Fornax, Draco, and Carina galaxies all have average $\mathrm{[Eu/Fe]}$ abundances of $\sim$0.4-0.5 dex \citep{letarte2010, aoki2007_umi, cohen2010, shetrone2003, norris2017, venn2012, cohen2009}. On the other hand, it has been observed that most surviving UFD galaxies have very low neutron-capture element abundances \citep{ji2019a}. However, as mentioned in the introduction to this paper, there are UFD galaxies that also exhibit an enhancement in $r$-process elements. Out of the nineteen UFD galaxies studied with high-resolution chemical abundance analyses to date, four ($\sim$20\%) have at least one star enhanced in $r$-process elements, specifically Reticulum~II, Tucana~III, Grus~II, and Tucana~V \citep{ji2016c, ji2023a, marshall2019, hansen2017, hansen2020, hansen2024}. Hence, while the enhanced $r$-process element abundance signature we see in Turranburra would point to it originating from a more massive DG, the UFD galaxy progenitor scenario is not excluded by this. Further, using the CaT metallicity from \cite{li2022}, the expected stellar mass from the stellar mass-metallicity relation from \cite{kirby2013} is $\log{(M_*/M_{\odot})}\sim 4.3$, placing Turranburra in the UFD galaxy mass regime. 

To further investigate the $r$-process element enhancement in streams originating from DGs, we plot the $\mathrm{[Eu II/Fe]}$ abundances of stars in classical DGs, stellar streams with DG progenitors, and UFD galaxies in Figure \ref{Fig:Eu_Fe}. The Eu abundances seen in the stellar streams, especially in Turranburra, are compatible with the enhancement that is seen in both the surviving classical DGs and UFD galaxies. However, as stated above, due to the low metallicity of Turranburra, it is more similar to the UFD galaxies. Furthermore, as can be seen in Figure \ref{Fig:Eu_Fe}, in the majority of the UFDs with $r$-process enhanced stars, the stars exhibit an Eu enhancement similar to what is seen in Turranburra and the other stream stars, while the extreme enhancement seen in Reticulum~II \citep{ji2016c} appears to be more rare. Given that the field of stellar streams is rapidly evolving, the likely detection of more streams in the future, particularly in the low-mass regime, will aid our understanding of systems of different masses. 

To gain a more complete view of the $r$-process signatures in Turranburra and Willka Yaku, we next explore the neutron-capture abundance patterns of the individual stars. Figure \ref{Fig:r_process_pattern} shows the neutron capture abundances for all the stream stars analyzed in this paper, excluding the RRL star Tur-3, compared with the abundances of the highly $r$-process enhanced, $r$-II star HD~222925 \citep{roederer2022}. As can be seen in \ref{Fig:r_process_pattern}, the heavy elements from Ba to Yb in the stream stars generally follow the pattern seen in HD~222925. A similar match is seen across $r$-process enhanced stars in the literature, which is known as the universal pattern \citep{roederer2022}. For the lighter elements Sr, Y, and Zr, Tur-1 and Tur-2 follow HD~222925, while WY-1, 2, and 3 are more enhanced in these elements relative to Eu than HD~222925. It has recently been shown by \cite{roederer2022} that if the light elements are scaled to Zr, instead of scaling the full pattern to Eu, a similar universal pattern emerges for Sr, Y, and Zr. This can also be expressed as a variation in the $\mathrm{[Sr/Eu]}$ ratios between the stars as was done in \cite{ji2019b}. The $\mathrm{[Sr/Eu]}$ ratio is directly relatable to the lanthanide fraction X$_{\rm La}$, which is the mass ratio between the heavy neutron-capture elements, La to Lu, and the total mass. A higher or lower X$_{\rm La}$ reflects changes in the ratio between the light and heavy $r$-process elements, and thus signals variation in the conditions of the $r$-process. The relative offsets in $\mathrm{[Sr/Eu]}$ between Turranburra and Willka Yaku therefore correspond to differences in their X$_{\rm La}$, confirming that the two streams experienced distinct $r$-process enrichment histories. As discussed in \cite{ji2019b}, neutron-star merger models with varying ejecta compositions can reproduce a broad range of lanthanide fractions, and the observed differences between Turranburra and Willka Yaku may reflect such diversity in the astrophysical $r$-process site. Alternatively, this could indicate contributions from additional sites that only produce Sr, Y, and Zr in Willka Yaku compared to Turranburra, such as the light element primary process or weak/limited $r$-process \citep{Travaglio2004,SiqueraMello2014,Hansen2018}. The clear difference seen between these two streams suggests that detailed $r$-process abundances can likely help with chemical tagging of stream stars.

\section{Summary \label{sec:summary}}

We have presented an abundance analysis of three stars each in the Willka Yaku and Turranburra stellar streams. For Willka Yaku we find a low metallicity dispersion ($\sim0.02$ dex) and abundances generally supporting the initial classification of GC for the stream progenitor \citep{li2022}. Further observations and analysis of more stars in Willka Yaku are needed to explore a possible spread in the light elements and low dispersion in the $\mathrm{[Eu/Fe]}$ abundances. For Turranburra, the overall abundance pattern and low metallicity suggest that Turranburra originated from a DG, potentially even a very low-mass system. Obtaining high-resolution spectra for abundance analysis of more stars in Turranburra, as well as other low-mass stellar streams, would help to confirm the UFD galaxy origin. Finally, both streams exhibit a mild enhancement in $r$-process elements, a feature they share with the majority of streams in the literature. Although both streams show an enhancement in $r$-process elements, the abundance pattern for the individual stars reveals differences in the ratio of light to heavy neutron-capture elements, indicating that the $r$-processes that enriched Turranburra and Willka Yaku were different.

\acknowledgments
This work is part of the ongoing \SSSSS program (\url{https://s5collab.github.io}).
\SSSSS includes data obtained with the Anglo-Australian Telescope in Australia. We acknowledge the traditional owners of the land on which the AAT stands, the Gamilaraay people, and pay our respects to elders past and present.


T.T.H acknowledges support from the Swedish Research Council (VR 2021-05556). 
A.P.J. and S.A.U. acknowledge support from the National Science Foundation grant AST-2206264.

G.E.M. acknowledges financial support from Natural Sciences and Engineering Research Council of Canada (NSERC) through grant RGPIN-2022-04794. G.E.M. acknowledges support from an Arts \& Science Postdoctoral Fellowship at the University of Toronto. The Dunlap Institute is funded through an endowment established by the David Dunlap family and the University of Toronto

This research made extensive use of the SIMBAD database operated at CDS, Straasburg, France \citep{wenger2000}, \href{https://arxiv.org/}{arXiv.org}, and NASA's Astrophysics Data System for bibliographic information.  

This work has made use of data from the European Space Agency (ESA) mission {\it Gaia} (\url{https://www.cosmos.esa.int/gaia}), processed by the {\it Gaia} Data Processing and Analysis Consortium (DPAC, \url{https://www.cosmos.esa.int/web/gaia/dpac/consortium}). Funding for the DPAC has been provided by national institutions, in particular, the institutions participating in the {\it Gaia} Multilateral Agreement.

NOIRLab IRAF is distributed by the Community Science and Data Center at NSF NOIRLab, which is managed by the Association of Universities for Research in Astronomy (AURA) under a cooperative agreement with the U.S. National Science Foundation.

\facility{Magellan:Clay}
\software{MOOG \citep{sneden1973,sobeck2011}, IRAF \citep{tody1986,tody1993, fitzpatrick2024}, ATLAS9 \citep{castelli2003}, linemake \citep{placco2021}, NumPy \citep{numpy, numpy1}, Matplotlib \citep{matplotlib}, AstroPy \citep{Astropy:13,Astropy:18, Astropy:22}, CarPy \citep{kelson2000, kelson2003}, 
SMHR \citep{casey2025}}

\appendix
\counterwithin{figure}{section}
\counterwithin{table}{section}

\section{}

\begin{deluxetable*}{crrrrrrrrrrrrrrrc}[hbp]
\tablecaption{ \label{tab:atomic_data_tur}Data for atomic lines used in analysis and individual line EW measurements for stars in Turranburra}
\tablehead{ \colhead{} & \colhead{} &\colhead{} & \multicolumn{4}{c}{Tur-1}  &\multicolumn{4}{c}{Tur-2} &\multicolumn{4}{c}{Tur-3}  \\
 \colhead{Species} & \colhead{$\lambda$} & \colhead{$\chi$} & \colhead{$\log{gf}$} & \colhead{EW} &  \colhead{$\sigma_{\rm EW}$} & \colhead{$\log{\epsilon}$} & \colhead{$\log{gf}$} & \colhead{EW} &  \colhead{$\sigma_{\rm EW}$} & \colhead{$\log{\epsilon}$} & \colhead{$\log{gf}$} & \colhead{EW} &  \colhead{$\sigma_{\rm EW}$} & \colhead{$\log{\epsilon}$}   & \colhead{ref}\\ 
 \colhead{} &\colhead{(\AA)} & \colhead{(eV)} & \colhead{} & \colhead{(m\AA)} & \colhead{(m\AA)} &\colhead{} &\colhead{}  & \colhead{(m\AA)} & \colhead{(m\AA)} &\colhead{} &\colhead{} & \colhead{(m\AA)} & \colhead{(m\AA)} &\colhead{} &\colhead{} & \colhead{}} 
 \startdata
 \ion{Na}{1}	&	5889	&	0.00	&	0.11	&	188.36	&	2.60	&	4.03	&	0.11	&	175.01	&	3.28	&	3.87	&	\nodata	&	\nodata	&	\nodata	&	\nodata	& 4 \\
\ion{Na}{1}	&	5895	&	0.00	&	$-$0.19	&	168.41	&	2.50	&	4.06	&	$-$0.19	&	151.45	&	2.99	&	3.85	&	$-$0.19	&	53.52	&	8.56	&	3.39	& 4 \\
\ion{Mg}{1}	&	3986	&	4.35	&	\nodata	&	\nodata	&	\nodata	&	\nodata	&	$-$1.06	&	40.44	&	4.44	&	5.49	&	\nodata	&	\nodata	&	\nodata	&	\nodata	 & 4 \\
\ion{Mg}{1}	&	4571	&	0.00	&	$-$5.62	&	68.67	&	1.76	&	5.46	&	$-$5.62	&	81.08	&	2.29	&	5.70	&	\nodata	&	\nodata	&	\nodata	&	\nodata	& 4\\
\ion{Mg}{1}	&	4702	&	4.35	&	$-$0.44	&	74.76	&	2.00	&	5.30	&	$-$0.44	&	72.30	&	2.32	&	5.27	&	\nodata	&	\nodata	&	\nodata	&	\nodata	& 4 \\
\ion{Mg}{1}	&	5172	&	2.71	&	$-$0.39	&	214.25	&	4.73	&	5.32	&	\nodata	&	\nodata	&	\nodata	&	\nodata	&	$-$0.39	&	150.39	&	8.35	&	5.41 & 4	\\

 \enddata 
 \tablerefs{ (1)\cite{masseron2014}; (2)\cite{sneden2014}; (3)\cite{caffau2008}; (4)\cite{kramida2019}; (5)\cite{ryabchikova2015}; (6)\cite{lawler1989}, using hfs from \cite{kurucz1995}; (7)\cite{lawler2013}; (8)\cite{wood2013}; (9)\cite{lawler2014}; (10)\cite{wood2014a}; (11)\cite{sobeck2007};(12)\cite{lawler2017}; (13)\cite{denhartog2011};(14)\cite{belmonte2017}; (15)\cite{denhartog2014}; (16)\cite{obrian1991}; (17)\cite{ruffoni2014}; (18)\cite{melendez2009}; (19)\cite{denhartog2019};(20)\cite{lawler2015}; (21)\cite{wood2014b}; (22)\cite{roederer2012}; (23)\cite{biemont2011};(24)\cite{hannaford1982}; (25)\cite{ljung2006};(26)\cite{mcwilliam1998}; (27)\cite{lawler2001a};(28)\cite{lawler2009}; (29)\cite{denhartog2003};(30)\cite{lawler2006}; (31)\cite{2001b}; (32)\cite{denhartog2006}; (33)\cite{sneden2009}; (34)\cite{yu2018}; (35)\cite{li2007}, using hfs from \cite{sneden2009}; (36)\cite{lawler2004}; (37)\cite{lawler2008b}; (38)\cite{lawler2011}}
\tablecomments{The complete version of this table is available online only. A short
version is shown here to illustrate its form and content.} 
\end{deluxetable*}

\begin{deluxetable*}{crrrrrrrrrrrrrrrc}
\tablecaption{ \label{tab:atomic_data_wy}Data for atomic lines used in analysis and individual line EW measurements for stars in Willka Yaku}
\tablehead{ \colhead{} & \colhead{} &\colhead{} & \multicolumn{4}{c}{WY-1}  &\multicolumn{4}{c}{WY-2} &\multicolumn{4}{c}{WY-3}  \\
 \colhead{Species} & \colhead{$\lambda$} & \colhead{$\chi$} & \colhead{$\log{gf}$} & \colhead{EW} &  \colhead{$\sigma_{\rm EW}$} & \colhead{$\log{\epsilon}$} & \colhead{$\log{gf}$} & \colhead{EW} &  \colhead{$\sigma_{\rm EW}$} & \colhead{$\log{\epsilon}$} & \colhead{$\log{gf}$} & \colhead{EW} &  \colhead{$\sigma_{\rm EW}$} & \colhead{$\log{\epsilon}$}  & \colhead{ref}\\ 
 \colhead{} &\colhead{(\AA)} & \colhead{(eV)} & \colhead{} & \colhead{(m\AA)} & \colhead{(m\AA)} &\colhead{} &\colhead{}  & \colhead{(m\AA)} & \colhead{(m\AA)} &\colhead{} &\colhead{} & \colhead{(m\AA)} & \colhead{(m\AA)} &\colhead{} &\colhead{} & \colhead{}} 
 \startdata
 \ion{Na}{1}	&	5889	&	0.00	&	0.11	&	240.42	&	2.98	&	4.04	&	0.11	&	226.35	&	6.01	&	4.45	&	0.11	&	211.48	&	4.03	&	4.40 & 4 \\
\ion{Na}{1}	&	5895	&	0.00	&	$-$0.19	&	215.35	&	2.59	&	4.09	&	$-$0.19	&	189.78	&	5.55	&	4.32	&	$-$0.19	&	180.06	&	3.92	&	4.29	& 4 \\
\ion{Mg}{1}	&	4167	&	4.35	&	$-$0.75	&	99.27	&	3.75	&	5.80	&	$-$0.75	&	69.83	&	5.44	&	5.57	&	$-$0.75	&	62.39	&	4.04	&	5.51	& 4 \\
\ion{Mg}{1}	&	4702	&	4.35	&	$-$0.44	&	120.99	&	2.10	&	5.67	&	$-$0.44	&	97.58	&	5.41	&	5.61	&	\nodata	&	\nodata	&	\nodata	&	\nodata	& 4 \\
\ion{Mg}{1}	&	5528	&	4.35	&	$-$0.50	&	123.58	&	3.15	&	5.71	&	$-$0.50	&	94.70	&	8.12	&	5.59	&	$-$0.50	&	87.58	&	3.75	&	5.54	& 4 \\
\ion{Mg}{1}	&	5711	&	4.34	&	$-$1.72	&	43.34	&	2.91	&	5.89	&	$-$1.72	&	28.48	&	6.30	&	5.83	&	\nodata	&	\nodata	&	\nodata	&	\nodata	& 4 \\

 \enddata 
 \tablerefs{ (1)\cite{masseron2014}; (2)\cite{sneden2014}; (3)\cite{caffau2008}; (4)\cite{kramida2019}; (5)\cite{ryabchikova2015}; (6)\cite{lawler1989}, using hfs from \cite{kurucz1995}; (7)\cite{lawler2013}; (8)\cite{wood2013}; (9)\cite{lawler2014}; (10)\cite{wood2014a}; (11)\cite{sobeck2007};(12)\cite{lawler2017}; (13)\cite{denhartog2011};(14)\cite{belmonte2017}; (15)\cite{denhartog2014}; (16)\cite{obrian1991}; (17)\cite{ruffoni2014}; (18)\cite{melendez2009}; (19)\cite{denhartog2019};(20)\cite{lawler2015}; (21)\cite{wood2014b}; (22)\cite{roederer2012}; (23)\cite{biemont2011};(24)\cite{hannaford1982}; (25)\cite{ljung2006};(26)\cite{mcwilliam1998}; (27)\cite{lawler2001a};(28)\cite{lawler2009}; (29)\cite{denhartog2003};(30)\cite{lawler2006}; (31)\cite{2001b}; (32)\cite{denhartog2006}; (33)\cite{sneden2009}; (34)\cite{yu2018}; (35)\cite{li2007}, using hfs from \cite{sneden2009}; (36)\cite{lawler2004}; (37)\cite{lawler2008b}; (38)\cite{lawler2011}}
\tablecomments{The complete version of this table is available online only. A short
version is shown here to illustrate its form and content.} 
\end{deluxetable*}

\setlength{\tabcolsep}{2pt}
\begin{deluxetable*}{ccrrrrrrrrrrrrrrrc}

\tablecaption{\label{tab:temperatures} Dereddened Magnitudes and Temperatures from Various Color Bands}
\tablehead{  \colhead{Name} & \colhead{$E(B-V)$} & \colhead{$G_0$} & \colhead{$BP_0$} & \colhead{$RP_0$} & \colhead{$K_0$} & \colhead{$T_{BP-RP}$} & \colhead{$T_{BP-G}$} & \colhead{$T_{G-RP}$} & \colhead{$T_{BP-K}$} &\colhead{$T_{RP-K}$} & \colhead{$T_{G-K}$} & \colhead{$\sigma$} & \colhead{$\sigma_{std}$} & \colhead{Avg.}\\
 \colhead{} & \colhead{} & \colhead{(mag)} &\colhead{(mag)}  & \colhead{(mag)} & \colhead{(mag)} & \colhead{(K)}  & \colhead{(K)}  & \colhead{(K)}  & \colhead{(K)}  & \colhead{(K)}  & \colhead{(K)}  & \colhead{(K)} & \colhead{(K)}  & \colhead{(K)} }
\startdata
\multirow{1}{*}{Tur-1}
& 0.06 & 16.49 & 16.94 & 15.88 & 14.48 & $4953\pm83$ & $4913\pm84$ & $4996\pm72$ & $4908\pm49$ & $4869\pm61$ & $4907\pm46$ & $\pm68$ & $\pm40$ & $4924\pm79$ \\
\hline
\multirow{1}{*}{Tur-2}
& 0.06 & 16.58 & 17.04 & 15.96 & 14.66 & $4916\pm83$  & $4888\pm84$ & $4945\pm72$ & $4981\pm49$ & $5036\pm61$ & $5010\pm46$ & $\pm68$ & $\pm52$ & $4963\pm85$ \\
\hline
\multirow{1}{*}{WY-1}
& 0.03 & 15.85  & 16.42  & 15.14  & 13.58  & $4575\pm83$  & $4557\pm84$  & $4590\pm72$  & $4586\pm49$  & $4607\pm61$  & $4605\pm46$  & $\pm68$ & $\pm17$ & $4587\pm70$  \\
\hline
\multirow{1}{*}{WY-2}
&  0.03 & 16.62  & 17.07  & 16.01  & 14.48  & $4968\pm83$  & $4952\pm84$  & $4985\pm72$  & $4789\pm49$  & $4646\pm61$  & $4740\pm46$ & $\pm68$ & $\pm129$ & $4847\pm146$  \\
\hline
\multirow{1}{*}{WY-3}
& 0.03 & 16.83 & 17.27  & 16.21  & 14.86  & $4958\pm83$  & $4931\pm84$  & $4987\pm72$  & $4943\pm49$  & $4928\pm61$  & $4946\pm46$  & $\pm68$ & $\pm20$ & $4949\pm70$ \\

\enddata
\tablecomments{$E(B-V)$ values from \cite{schlafly2011}. $G_0$, $BP_0$, $RP_0$, and $K_0$ are dereddened photometry values for each star. T($BP-RP$), T($BP-G$), T($BP-K$),T($RP-K$), T($G-K$) are the temperatures and associated uncertainties derived from the specified color band accounting for the metallicity and uncertainties from \cite{mucciarelli2021}. $\sigma$ is the mean error across all color bands. $\sigma_{std}$ is the standard deviation of the temperatures across all color bands. Avg. is the average $T_{\rm eff}$ and total uncertainty.}
\end{deluxetable*}

\begin{deluxetable*}{crrrrrrrrrrrrrrrrrc}

\tablecaption{\label{tab:tur_uncertainties}Turranburra Uncertainties}
\tablehead{
\colhead{} & \multicolumn{6}{c}{Tur-1} &\multicolumn{6}{c}{Tur-2} &\multicolumn{6}{c}{Tur-3}  \\
\colhead{Species} & \colhead{N} &  \colhead{$\Delta_{T_{\rm eff}}$} & \colhead{$\Delta_{\log g}$} & \colhead{$\Delta_{\xi}$} & \colhead{$\Delta_\mathrm{[Fe/H]}$}& \colhead{$s_X$} &  \colhead{N} &  \colhead{$\Delta_{T_{\rm eff}}$} & \colhead{$\Delta_{\log g}$} & \colhead{$\Delta_{\xi}$} & \colhead{$\Delta_\mathrm{[Fe/H]}$}& \colhead{$s_X$} &  \colhead{N} & \colhead{$\Delta_{T_{\rm eff}}$} & \colhead{$\Delta_{\log g}$} & \colhead{$\Delta_{\xi}$} & \colhead{$\Delta_\mathrm{[Fe/H]}$}& \colhead{$s_X$}  \\
\colhead{} & \colhead{} & \colhead{(dex)} & \colhead{(dex)} & \colhead{(dex)} & \colhead{(dex)} & \colhead{(dex)} & \colhead{} & \colhead{(dex)} & \colhead{(dex)} & \colhead{(dex)} & \colhead{(dex)} & \colhead{(dex)} & \colhead{} & \colhead{(dex)} & \colhead{(dex)} & \colhead{(dex)} & \colhead{(dex)} & \colhead{(dex)} }
\startdata
\ion{CH}{0}	&	3	&	0.20	&	$-$0.10	&	0.01	&	0.07	&	0.03	&	3	&	0.22	&	$-$0.06	&	0.01	&	0.05	&	0.00	&	\nodata	&	\nodata	&	\nodata	&	\nodata	&	\nodata	&	\nodata	\\
\ion{Na}{1}	&	2	&	0.16	&	$-$0.05	&	$-$0.12	&	$-$0.02	&	0.00	&	2	&	0.15	&	$-$0.04	&	$-$0.11	&	$-$0.01	&	0.00	&	1	&	0.30	&	$-$0.02	&	$-$0.01	&	0.01	&	0.00	\\
\ion{Mg}{1}	&	5	&	0.12	&	$-$0.05	&	$-$0.05	&	0.00	&	0.10	&	5	&	0.09	&	$-$0.01	&	$-$0.02	&	0.00	&	0.16	&	3	&	0.24	&	$-$0.01	&	$-$0.08	&	0.00	&	0.22	\\
\ion{Al}{1}	&	2	&	0.25	&	$-$0.17	&	$-$0.12	&	0.07	&	0.36	&	2	&	0.47	&	$-$0.16	&	$-$0.12	&	0.03	&	0.30	&	1	&	0.32	&	$-$0.01	&	$-$0.02	&	0.00	&	0.30	\\
\ion{Si}{1}	&	2	&	0.32	&	$-$0.10	&	$-$0.10	&	0.05	&	0.54	&	1	&	0.26	&	$-$0.09	&	0.00	&	0.03	&	0.00	&	1	&	0.32	&	$-$0.01	&	$-$0.04	&	0.01	&	0.00	\\
\ion{K}{1}	&	1	&	0.09	&	$-$0.01	&	$-$0.02	&	$-$0.01	&	0.00	&	2	&	0.11	&	0.00	&	$-$0.04	&	0.00	&	0.08	&	\nodata	&	\nodata	&	\nodata	&	\nodata	&	\nodata	&	\nodata	\\
\ion{Ca}{1}	&	15	&	0.09	&	$-$0.01	&	$-$0.03	&	$-$0.01	&	0.12	&	17	&	0.09	&	$-$0.01	&	$-$0.04	&	0.00	&	0.10	&	3	&	0.28	&	$-$0.02	&	0.00	&	0.01	&	0.10	\\
\ion{Sc}{2}	&	8	&	$-$0.01	&	0.07	&	$-$0.03	&	0.01	&	0.12	&	7	&	$-$0.01	&	0.08	&	$-$0.05	&	0.00	&	0.10	&	3	&	0.13	&	0.06	&	$-$0.01	&	0.00	&	0.10	\\
\ion{Ti}{1}	&	19	&	0.14	&	$-$0.01	&	$-$0.02	&	$-$0.01	&	0.11	&	15	&	0.14	&	$-$0.01	&	$-$0.03	&	0.00	&	0.15	&	\nodata	&	\nodata	&	\nodata	&	\nodata	&	\nodata	&	\nodata	\\
\ion{Ti}{2}	&	21	&	0.04	&	0.06	&	$-$0.02	&	0.02	&	0.12	&	28	&	0.05	&	0.08	&	$-$0.06	&	0.02	&	0.20	&	14	&	0.17	&	0.07	&	$-$0.05	&	0.00	&	0.17	\\
\ion{V}{1}	&	2	&	0.08	&	0.01	&	0.00	&	$-$0.02	&	0.00	&	2	&	0.12	&	0.03	&	0.01	&	0.00	&	0.00	&	\nodata	&	\nodata	&	\nodata	&	\nodata	&	\nodata	&	\nodata	\\
\ion{V}{2}	&	2	&	$-$0.01	&	0.06	&	0.01	&	0.03	&	0.00	&	2	&	$-$0.01	&	0.08	&	0.01	&	0.01	&	0.00	&	\nodata	&	\nodata	&	\nodata	&	\nodata	&	\nodata	&	\nodata	\\
\ion{Cr}{1}	&	12	&	0.13	&	$-$0.01	&	$-$0.02	&	$-$0.01	&	0.18	&	6	&	0.14	&	$-$0.01	&	$-$0.03	&	0.00	&	0.15	&	\nodata	&	\nodata	&	\nodata	&	\nodata	&	\nodata	&	\nodata	\\
\ion{Cr}{2}	&	4	&	$-$0.02	&	0.07	&	$-$0.01	&	0.01	&	0.06	&	3	&	$-$0.01	&	0.08	&	$-$0.01	&	0.01	&	0.08	&	1	&	0.06	&	0.07	&	$-$0.01	&	0.00	&	0.00	\\
\ion{Mn}{1}	&	7	&	0.11	&	$-$0.01	&	$-$0.04	&	$-$0.02	&	0.25	&	6	&	0.15	&	0.00	&	$-$0.05	&	$-$0.01	&	0.12	&	\nodata	&	\nodata	&	\nodata	&	\nodata	&	\nodata	&	\nodata	\\
\ion{Fe}{1}	&	75	&	0.13	&	$-$0.01	&	$-$0.06	&	$-$0.01	&	0.12	&	63	&	0.14	&	$-$0.01	&	$-$0.07	&	0.00	&	0.13	&	12	&	0.36	&	$-$0.01	&	$-$0.04	&	0.01	&	0.15	\\
\ion{Fe}{2}	&	14	&	0.00	&	0.06	&	$-$0.03	&	0.02	&	0.10	&	14	&	0.01	&	0.08	&	$-$0.04	&	0.02	&	0.09	&	9	&	0.09	&	0.07	&	$-$0.03	&	0.00	&	0.09	\\
\ion{Co}{1}	&	6	&	0.12	&	0.00	&	$-$0.02	&	$-$0.02	&	0.25	&	6	&	0.15	&	0.04	&	$-$0.01	&	0.00	&	0.00	&	\nodata	&	\nodata	&	\nodata	&	\nodata	&	\nodata	&	\nodata	\\
\ion{Ni}{1}	&	12	&	0.11	&	0.00	&	$-$0.01	&	0.00	&	0.11	&	8	&	0.11	&	0.00	&	$-$0.02	&	0.01	&	0.15	&	1	&	0.34	&	$-$0.01	&	0.01	&	0.01	&	0.00	\\
\ion{Cu}{1}	&	\nodata	&	\nodata	&	\nodata	&	\nodata	&	\nodata	&	\nodata	&	\nodata	&	\nodata	&	\nodata	&	\nodata	&	\nodata	&	\nodata	&	\nodata	&	\nodata	&	\nodata	&	\nodata	&	\nodata	&	\nodata	\\
\ion{Zn}{1}	&	2	&	0.04	&	0.03	&	$-$0.01	&	0.01	&	0.00	&	2	&	0.05	&	0.05	&	$-$0.01	&	0.01	&	0.16	&	\nodata	&	\nodata	&	\nodata	&	\nodata	&	\nodata	&	\nodata	\\
\ion{Sr}{2}	&	2	&	0.00	&	0.09	&	$-$0.20	&	$-$0.01	&	0.10	&	2	&	0.05	&	0.06	&	$-$0.21	&	0.00	&	0.00	&	2	&	0.29	&	0.06	&	$-$0.10	&	0.00	&	0.04	\\
\ion{Y}{2}	&	2	&	0.05	&	0.05	&	0.01	&	0.03	&	0.00	&	5	&	0.04	&	0.08	&	0.00	&	0.02	&	0.00	&	\nodata	&	\nodata	&	\nodata	&	\nodata	&	\nodata	&	\nodata	\\
\ion{Zr}{2}	&	1	&	$-$0.05	&	0.10	&	0.01	&	0.01	&	0.00	&	1	&	$-$0.07	&	0.10	&	0.00	&	$-$0.01	&	0.00	&	\nodata	&	\nodata	&	\nodata	&	\nodata	&	\nodata	&	\nodata	\\
\ion{Ba}{2}	&	5	&	0.04	&	0.05	&	$-$0.04	&	0.01	&	0.14	&	5	&	0.06	&	0.08	&	$-$0.07	&	0.01	&	0.11	&	1	&	0.34	&	0.00	&	0.01	&	0.00	&	0.10	\\
\ion{La}{2}	&	2	&	0.03	&	0.06	&	$-$0.01	&	0.02	&	0.00	&	2	&	0.05	&	0.07	&	0.00	&	0.02	&	0.00	&	\nodata	&	\nodata	&	\nodata	&	\nodata	&	\nodata	&	\nodata	\\
\ion{Ce}{2}	&	\nodata	&	\nodata	&	\nodata	&	\nodata	&	\nodata	&	\nodata	&	\nodata	&	\nodata	&	\nodata	&	\nodata	&	\nodata	&	\nodata	&	\nodata	&	\nodata	&	\nodata	&	\nodata	&	\nodata	&	\nodata	\\
\ion{Pr}{2}	&	\nodata	&	\nodata	&	\nodata	&	\nodata	&	\nodata	&	\nodata	&	1	&	0.13	&	0.10	&	$-$0.01	&	0.03	&	0.00	&	\nodata	&	\nodata	&	\nodata	&	\nodata	&	\nodata	&	\nodata	\\
\ion{Nd}{2}	&	1	&	0.02	&	0.05	&	0.00	&	0.03	&	0.00	&	2	&	0.04	&	0.08	&	0.05	&	0.08	&	0.00	&	\nodata	&	\nodata	&	\nodata	&	\nodata	&	\nodata	&	\nodata	\\
\ion{Sm}{2}	&	\nodata	&	\nodata	&	\nodata	&	\nodata	&	\nodata	&	\nodata	&	\nodata	&	\nodata	&	\nodata	&	\nodata	&	\nodata	&	\nodata	&	\nodata	&	\nodata	&	\nodata	&	\nodata	&	\nodata	&	\nodata	\\
\ion{Eu}{2}	&	2	&	0.03	&	0.06	&	0.01	&	0.02	&	0.00	&	3	&	0.08	&	0.08	&	0.00	&	0.02	&	0.00	&	\nodata	&	\nodata	&	\nodata	&	\nodata	&	\nodata	&	\nodata	\\
\ion{Gd}{2}	&	\nodata	&	\nodata	&	\nodata	&	\nodata	&	\nodata	&	\nodata	&	\nodata	&	\nodata	&	\nodata	&	\nodata	&	\nodata	&	\nodata	&	\nodata	&	\nodata	&	\nodata	&	\nodata	&	\nodata	&	\nodata	\\
\ion{Dy}{2}	&	1	&	0.03	&	0.06	&	$-$0.02	&	0.02	&	0.00	&	1	&	0.06	&	0.08	&	$-$0.02	&	0.03	&	0.00	&	\nodata	&	\nodata	&	\nodata	&	\nodata	&	\nodata	&	\nodata	\\
\ion{Ho}{2}	&	\nodata	&	\nodata	&	\nodata	&	\nodata	&	\nodata	&	\nodata	&	\nodata	&	\nodata	&	\nodata	&	\nodata	&	\nodata	&	\nodata	&	\nodata	&	\nodata	&	\nodata	&	\nodata	&	\nodata	&	\nodata	\\
\ion{Er}{2}	&	2	&	$-$0.06	&	0.10	&	$-$0.03	&	0.01	&	0.00	&	2	&	0.02	&	0.09	&	$-$0.05	&	0.01	&	0.00	&	\nodata	&	\nodata	&	\nodata	&	\nodata	&	\nodata	&	\nodata	\\
\ion{Yb}{2}	&	1	&	0.07	&	0.07	&	$-$0.01	&	0.02	&	0.00	&	\nodata	&	\nodata	&	\nodata	&	\nodata	&	\nodata	&	\nodata	&	\nodata	&	\nodata	&	\nodata	&	\nodata	&	\nodata	&	\nodata	\\

\enddata
\end{deluxetable*}

\begin{deluxetable*}{crrrrrrrrrrrrrrrrrc}

\tablecaption{\label{tab:wy_uncertainties}Willka Yaku Uncertainties}
\tablehead{
\colhead{} & \multicolumn{6}{c}{WY-1 } &\multicolumn{6}{c}{WY-2} &\multicolumn{6}{c}{WY-3}   \\
\colhead{Species} & \colhead{N} &  \colhead{$\Delta_{T_{\rm eff}}$} & \colhead{$\Delta_{\log g}$} & \colhead{$\Delta_{\xi}$} & \colhead{$\Delta_\mathrm{[Fe/H]}$}& \colhead{$s_X$} &  \colhead{N} &  \colhead{$\Delta_{T_{\rm eff}}$} & \colhead{$\Delta_{\log g}$} & \colhead{$\Delta_{\xi}$} & \colhead{$\Delta_\mathrm{[Fe/H]}$}& \colhead{$s_X$} &  \colhead{N} & \colhead{$\Delta_{T_{\rm eff}}$} & \colhead{$\Delta_{\log g}$} & \colhead{$\Delta_{\xi}$} & \colhead{$\Delta_\mathrm{[Fe/H]}$}& \colhead{$s_X$} \\
\colhead{} & \colhead{} & \colhead{(dex)} & \colhead{(dex)} & \colhead{(dex)} & \colhead{(dex)} & \colhead{(dex)} & \colhead{} & \colhead{(dex)} & \colhead{(dex)} & \colhead{(dex)} & \colhead{(dex)} & \colhead{(dex)} & \colhead{} & \colhead{(dex)} & \colhead{(dex)} & \colhead{(dex)} & \colhead{(dex)} & \colhead{(dex)} }
\startdata
\ion{CH} {0}	&	3	&	0.09	&	0.00	&	0.01	&	0.09	&	0.03	&	2	&	0.22	&	$-$0.06	&	0.02	&	0.09	&	0.09	&	2	&	0.20	&	$-$0.09	&	0.02	&	0.02	&	0.02	\\
\ion{CN}{0}	&	1	&	0.23	&	0.00	&	0.01	&	0.14	&	0.00	&	\nodata	&	\nodata	&	\nodata	&	\nodata	&	\nodata	&	\nodata	&	\nodata	&	\nodata	&	\nodata	&	\nodata	&	\nodata	&	\nodata	\\
\ion{O}{0} 	&	1	&	0.04	&	0.08	&	0.00	&	0.06	&	0.00	&	\nodata	&	\nodata	&	\nodata	&	\nodata	&	\nodata	&	\nodata	&	\nodata	&	\nodata	&	\nodata	&	\nodata	&	\nodata	&	\nodata	\\
\ion{Na}{1} 	&	3	&	0.16	&	$-$0.04	&	$-$0.07	&	$-$0.01	&	0.09	&	2	&	0.16	&	$-$0.07	&	$-$0.11	&	$-$0.05	&	0.00	&	2	&	0.16	&	$-$0.04	&	$-$0.12	&	$-$0.02	&	0.01	\\
\ion{Mg}{1} 	&	4	&	0.10	&	$-$0.03	&	$-$0.05	&	$-$0.01	&	0.09	&	9	&	0.10	&	$-$0.04	&	$-$0.04	&	$-$0.02	&	0.06	&	6	&	0.12	&	$-$0.03	&	$-$0.04	&	0.00	&	0.13	\\
\ion{Al}{1} 	&	\nodata	&	\nodata	&	\nodata	&	\nodata	&	\nodata	&	\nodata	&	1	&	1.10	&	$-$0.22	&	$-$0.16	&	0.18	&	0.30	&	1	&	0.36	&	$-$0.25	&	$-$0.16	&	0.01	&	0.30	\\
\ion{Si}{1} 	&	4	&	0.09	&	$-$0.02	&	$-$0.04	&	0.00	&	0.17	&	1	&	0.15	&	$-$0.06	&	$-$0.06	&	$-$0.01	&	0.00	&	1	&	0.14	&	$-$0.06	&	$-$0.05	&	0.00	&	0.00	\\
\ion{K}{1} 	&	2	&	0.16	&	$-$0.02	&	$-$0.08	&	$-$0.02	&	0.00	&	2	&	0.11	&	$-$0.03	&	$-$0.06	&	$-$0.03	&	0.13	&	2	&	0.11	&	0.00	&	$-$0.06	&	0.00	&	0.00	\\
\ion{Ca}{1} 	&	20	&	0.11	&	$-$0.03	&	$-$0.05	&	$-$0.01	&	0.15	&	15	&	0.09	&	$-$0.03	&	$-$0.05	&	$-$0.02	&	0.16	&	12	&	0.09	&	$-$0.01	&	$-$0.04	&	0.00	&	0.12	\\
\ion{Sc}{2} 	&	6	&	$-$0.01	&	0.05	&	$-$0.05	&	0.03	&	0.12	&	6	&	0.02	&	0.07	&	$-$0.06	&	0.01	&	0.13	&	7	&	$-$0.01	&	0.06	&	$-$0.07	&	$-$0.02	&	0.14	\\
\ion{Ti}{1} 	&	28	&	0.17	&	$-$0.03	&	$-$0.04	&	$-$0.03	&	0.11	&	14	&	0.13	&	$-$0.02	&	$-$0.02	&	$-$0.01	&	0.08	&	15	&	0.15	&	$-$0.01	&	$-$0.03	&	0.00	&	0.12	\\
\ion{Ti}{2} 	&	33	&	0.02	&	0.08	&	$-$0.07	&	0.05	&	0.16	&	25	&	0.05	&	0.07	&	$-$0.07	&	0.02	&	0.10	&	32	&	0.04	&	0.07	&	$-$0.07	&	0.00	&	0.11	\\
\ion{V}{1} 	&	3	&	0.17	&	$-$0.04	&	0.00	&	$-$0.03	&	0.00	&	2	&	0.13	&	0.01	&	0.00	&	0.00	&	0.00	&	2	&	0.14	&	0.00	&	0.00	&	0.00	&	0.00	\\
\ion{V}{2} 	&	2	&	$-$0.05	&	0.09	&	$-$0.01	&	0.05	&	0.00	&	2	&	0.01	&	0.08	&	0.00	&	0.03	&	0.00	&	2	&	$-$0.02	&	0.07	&	0.01	&	0.00	&	0.00	\\
\ion{Cr}{1} 	&	10	&	0.17	&	$-$0.03	&	$-$0.05	&	$-$0.03	&	0.16	&	5	&	0.13	&	$-$0.02	&	$-$0.02	&	$-$0.01	&	0.12	&	8	&	0.14	&	$-$0.01	&	$-$0.02	&	0.00	&	0.12	\\
\ion{Cr}{2} 	&	4	&	$-$0.03	&	0.08	&	$-$0.01	&	0.03	&	0.07	&	3	&	$-$0.01	&	0.07	&	$-$0.01	&	0.01	&	0.02	&	3	&	$-$0.02	&	0.07	&	$-$0.01	&	0.00	&	0.00	\\
\ion{Mn}{1} 	&	7	&	0.11	&	$-$0.03	&	$-$0.01	&	$-$0.03	&	0.11	&	5	&	0.10	&	$-$0.02	&	$-$0.02	&	$-$0.01	&	0.10	&	4	&	0.12	&	$-$0.02	&	$-$0.05	&	$-$0.01	&	0.13	\\
\ion{Fe}{1} 	&	71	&	0.16	&	$-$0.02	&	$-$0.07	&	$-$0.02	&	0.11	&	55	&	0.13	&	$-$0.03	&	$-$0.06	&	$-$0.03	&	0.05	&	66	&	0.14	&	$-$0.01	&	$-$0.07	&	$-$0.01	&	0.11	\\
\ion{Fe}{2} 	&	14	&	$-$0.01	&	0.09	&	$-$0.04	&	0.05	&	0.08	&	10	&	0.01	&	0.07	&	$-$0.04	&	0.02	&	0.06	&	19	&	0.00	&	0.07	&	$-$0.05	&	0.00	&	0.05	\\
\ion{Co}{1} 	&	3	&	0.14	&	$-$0.02	&	$-$0.06	&	$-$0.05	&	0.17	&	2	&	0.09	&	$-$0.03	&	$-$0.05	&	$-$0.08	&	0.00	&	2	&	0.14	&	$-$0.01	&	$-$0.04	&	0.00	&	0.00	\\
\ion{Ni}{1} 	&	18	&	0.13	&	$-$0.01	&	$-$0.02	&	$-$0.01	&	0.16	&	6	&	0.12	&	$-$0.02	&	$-$0.02	&	$-$0.01	&	0.00	&	13	&	0.11	&	$-$0.01	&	$-$0.02	&	0.00	&	0.11	\\
\ion{Cu}{1} 	&	1	&	0.15	&	$-$0.02	&	$-$0.01	&	$-$0.02	&	0.00	&	\nodata	&	\nodata	&	\nodata	&	\nodata	&	\nodata	&	\nodata	&	\nodata	&	\nodata	&	\nodata	&	\nodata	&	\nodata	&	\nodata	\\
\ion{Zn}{1} 	&	2	&	0.02	&	0.05	&	$-$0.01	&	0.03	&	0.00	&	2	&	0.06	&	0.03	&	$-$0.02	&	0.01	&	0.07	&	2	&	0.06	&	0.02	&	$-$0.01	&	0.00	&	0.00	\\
\ion{Sr}{2} 	&	2	&	0.07	&	0.08	&	$-$0.10	&	0.06	&	0.12	&	2	&	0.05	&	0.06	&	$-$0.15	&	0.02	&	0.00	&	2	&	0.07	&	0.06	&	$-$0.19	&	$-$0.02	&	0.00	\\
\ion{Y}{2} 	&	3	&	$-$0.01	&	0.08	&	$-$0.02	&	0.05	&	0.00	&	3	&	0.04	&	0.06	&	$-$0.01	&	0.02	&	0.00	&	3	&	0.04	&	0.06	&	$-$0.01	&	0.00	&	0.05	\\
\ion{Zr}{2} 	&	1	&	$-$0.08	&	0.10	&	$-$0.01	&	0.01	&	0.00	&	1	&	0.00	&	0.08	&	0.00	&	0.01	&	0.00	&	1	&	$-$0.01	&	0.07	&	$-$0.02	&	$-$0.01	&	0.00	\\
\ion{Ba}{2} 	&	5	&	0.02	&	0.08	&	$-$0.09	&	0.05	&	0.15	&	5	&	0.06	&	0.05	&	$-$0.09	&	0.01	&	0.22	&	5	&	0.06	&	0.06	&	$-$0.08	&	$-$0.01	&	0.13	\\
\ion{La}{2} 	&	4	&	0.00	&	0.08	&	0.00	&	0.05	&	0.00	&	3	&	0.04	&	0.07	&	0.00	&	0.03	&	0.07	&	3	&	0.03	&	0.06	&	0.00	&	0.00	&	0.00	\\
\ion{Ce}{2} 	&	2	&	0.03	&	0.10	&	$-$0.02	&	0.03	&	0.00	&	1	&	0.07	&	0.08	&	0.00	&	0.03	&	0.00	&	\nodata	&	\nodata	&	\nodata	&	\nodata	&	\nodata	&	\nodata	\\
\ion{Pr}{2} 	&	1	&	0.06	&	0.08	&	0.00	&	0.05	&	0.00	&	1	&	0.11	&	0.08	&	0.00	&	0.04	&	0.00	&	1	&	0.12	&	0.06	&	$-$0.02	&	0.00	&	0.00	\\
\ion{Nd}{2} 	&	3	&	0.01	&	0.09	&	$-$0.02	&	0.06	&	0.11	&	2	&	0.06	&	0.09	&	0.02	&	0.05	&	0.00	&	1	&	0.03	&	0.06	&	$-$0.02	&	0.00	&	0.00	\\
\ion{Sm}{2} 	&	1	&	$-$0.02	&	0.08	&	0.01	&	0.05	&	0.00	&	1	&	0.03	&	0.07	&	0.01	&	0.04	&	0.00	&	\nodata	&	\nodata	&	\nodata	&	\nodata	&	\nodata	&	\nodata	\\
\ion{Eu}{2} 	&	2	&	0.02	&	0.09	&	0.01	&	0.06	&	0.00	&	2	&	0.06	&	0.08	&	0.00	&	0.04	&	0.00	&	2	&	0.04	&	0.06	&	0.00	&	0.00	&	0.00	\\
\ion{Gd}{2} 	&	1	&	$-$0.03	&	0.11	&	0.01	&	0.07	&	0.00	&	1	&	0.00	&	0.00	&	0.00	&	0.00	&	0.00	&	\nodata	&	\nodata	&	\nodata	&	\nodata	&	\nodata	&	\nodata	\\
\ion{Dy}{2} 	&	2	&	0.01	&	0.08	&	$-$0.03	&	0.07	&	0.13	&	1	&	0.02	&	0.05	&	$-$0.03	&	0.02	&	0.00	&	1	&	0.07	&	0.06	&	$-$0.06	&	0.00	&	0.00	\\
\ion{Ho}{2} 	&	2	&	0.10	&	0.04	&	$-$0.01	&	0.04	&	0.16	&	\nodata	&	\nodata	&	\nodata	&	\nodata	&	\nodata	&	\nodata	&	2	&	0.18	&	0.04	&	$-$0.03	&	0.00	&	0.00	\\
\ion{Er}{2} 	&	2	&	$-$0.10	&	0.14	&	0.02	&	0.06	&	0.00	&	\nodata	&	\nodata	&	\nodata	&	\nodata	&	\nodata	&	\nodata	&	2	&	0.02	&	0.07	&	$-$0.03	&	$-$0.01	&	0.00	\\
\ion{Yb}{2} 	&	\nodata	&	\nodata	&	\nodata	&	\nodata	&	\nodata	&	\nodata	&	\nodata	&	\nodata	&	\nodata	&	\nodata	&	\nodata	&	\nodata	&	\nodata	&	\nodata	&	\nodata	&	\nodata	&	\nodata	&	\nodata	\\

\enddata
\end{deluxetable*}


\begin{thebibliography}{}
\bibitem[Amorisco(2015)]{amorisco2015} Amorisco, N.~C.\ 2015, \mnras, 450, 1, 575. doi:10.1093/mnras/stv648
\bibitem[Aoki et al.(2007)]{aoki2007} Aoki, W., Beers, T.~C., Christlieb, N., et al.\ 2007, \apj, 655, 492. doi:10.1086/509817
\bibitem[Aoki et al.(2007)]{aoki2007_umi} Aoki, W., Honda, S., Sadakane, K., et al.\ 2007, \pasj, 59, L15. doi:10.1093/pasj/59.3.L15
\bibitem[Asplund et al.(2009)]{asplund2009} Asplund, M., Grevesse, N., Sauval, A.~J., et al.\ 2009, \araa, 47, 481. doi:10.1146/annurev.astro.46.060407.145222
\bibitem[Astropy Collaboration et al.(2022)]{Astropy:22} Astropy Collaboration, Price-Whelan, A.~M., Lim, P.~L., et al.\ 2022, \apj, 935, 2, 167. doi:10.3847/1538-4357/ac7c74
\bibitem[Astropy Collaboration et al.(2013)]{Astropy:13} Astropy Collaboration, Robitaille, T.~P., Tollerud, E.~J., et al.\ 2013, \aap, 558, A33
\bibitem[Astropy Collaboration et al.(2018)]{Astropy:18} Astropy Collaboration, Price-Whelan, A.~M., Sip{\H{o}}cz, B.~M., et al.\ 2018, \aj, 156, 123
\bibitem[Atzberger et al.(2024)]{atzberger2024} Atzberger, K.~R., Usman, S.~A., Ji, A.~P., et al.\ 2024, arXiv:2410.17312. doi:10.48550/arXiv.2410.17312
\bibitem[Bechtol et al.(2015)]{bechtol2015} Bechtol, K., Drlica-Wagner, A., Balbinot, E., et al.\ 2015, \apj, 807, 50. doi:10.1088/0004-637X/807/1/50
\bibitem[Beers \& Christlieb(2005)]{beers2005} Beers, T.~C. \& Christlieb, N.\ 2005, \araa, 43, 531. doi:10.1146/annurev.astro.42.053102.134057
\bibitem[Belmonte et al.(2017)]{belmonte2017} Belmonte, M.~T., Pickering, J.~C., Ruffoni, M.~P., et al.\ 2017, \apj, 848, 125. doi:10.3847/1538-4357/aa8cd3
\bibitem[Bernstein et al.(2003)]{bernstein2003} Bernstein, R., Shectman, S.~A., Gunnels, S.~M., et al.\ 2003, \procspie, 4841, 1694. doi:10.1117/12.461502
\bibitem[Bi{\'e}mont et al.(2011)]{biemont2011} Bi{\'e}mont, {\'E}., Blagoev, K., Engstr{\"o}m, L., et al.\ 2011, \mnras, 414, 4, 3350. doi:10.1111/j.1365-2966.2011.18637.x
\bibitem[Bonaca \& Price-Whelan(2025)]{bonaca2025} Bonaca, A. \& Price-Whelan, A.~M.\ 2025, \nar, 100, 101713. doi:10.1016/j.newar.2024.101713
\bibitem[Bonaca et al.(2021)]{bonaca2021} Bonaca, A., Naidu, R.~P., Conroy, C., et al.\ 2021, \apjl, 909, 2, L26. doi:10.3847/2041-8213/abeaa9
\bibitem[Bland-Hawthorn et al.(2010)]{bland-hawthorn2010} Bland-Hawthorn, J., Krumholz, M.~R., \& Freeman, K.\ 2010, \apj, 713, 1, 166. doi:10.1088/0004-637X/713/1/166
\bibitem[Caffau et al.(2008)]{caffau2008} Caffau, E., Ludwig, H.-G., Steffen, M., et al.\ 2008, \aap, 488, 3, 1031. doi:10.1051/0004-6361:200809885
\bibitem[Carretta(2019)]{carretta2019} Carretta, E.\ 2019, \aap, 624, A24. doi:10.1051/0004-6361/201935110
\bibitem[Carretta et al.(2010)]{carretta2010} Carretta, E., Bragaglia, A., Gratton, R.~G., et al.\ 2010, \aap, 516, A55. doi:10.1051/0004-6361/200913451
\bibitem[Carretta et al.(2009)]{carretta2009} Carretta, E., Bragaglia, A., Gratton, R.~G., et al.\ 2009, \aap, 505, 1, 117. doi:10.1051/0004-6361/200912096
\bibitem[Casey et al.(2025)]{casey2025} Casey, A.~R., Ji, A., \& Holmbeck, E.\ 2025, Astrophysics Source Code Library, smhr: Automatic curve-of-growth analyses of high-resolution stellar spectra. ascl:2502.025
\bibitem[Castelli \& Kurucz(2003)]{castelli2003} Castelli, F., \& Kurucz,
R.~L.\ 2003, in IAU Symp. 210, Modelling of Stellar Atmospheres(Cambridge: Cambridge Univ. Press),A20. 
\bibitem[Cohen \& Huang(2009)]{cohen2009} Cohen, J.~G. \& Huang, W.\ 2009, \apj, 701, 1053. doi:10.1088/0004-637X/701/2/1053
\bibitem[Christlieb et al.(2004)]{christlieb2004} Christlieb, N., Beers, T.~C., Barklem, P.~S., et al.\ 2004, \aap, 428, 1027. doi:10.1051/0004-6361:20041536
\bibitem[Cohen \& Huang(2010)]{cohen2010} Cohen, J.~G. \& Huang, W.\ 2010, \apj, 719, 931. doi:10.1088/0004-637X/719/1/931
\bibitem[Cowan et al.(2021)]{cowan2021} Cowan, J.~J., Sneden, C., Lawler, J.~E., et al.\ 2021, Reviews of Modern Physics, 93, 015002. doi:10.1103/RevModPhys.93.015002
\bibitem[Cutri et al. (2003)]{cutri2003} Cutri, R. M., Skrutskie, M.F., van Dyk, S., et al. 2003, VizieR Online Data Catalog, 2246
\bibitem[DES Collaboration(2005)]{DES} The Dark Energy Survey Collaboration\ 2005, astro-ph/0510346. doi:10.48550/arXiv.astro-ph/0510346
\bibitem[Den Hartog et al.(2019)]{denhartog2019} Den Hartog, E.~A., Lawler, J.~E., Sneden, C., et al.\ 2019, \apjs, 243, 33. doi:10.3847/1538-4365/ab322e
\bibitem[Den Hartog et al.(2014)]{denhartog2014} Den Hartog, E.~A., Ruffoni, M.~P., Lawler, J.~E., et al.\ 2014, \apjs, 215, 23. doi:10.1088/0067-0049/215/2/23
\bibitem[Den Hartog et al.(2011)]{denhartog2011} Den Hartog, E.~A., Lawler, J.~E., Sobeck, J.~S., et al.\ 2011, \apjs, 194, 35. doi:10.1088/0067-0049/194/2/35
\bibitem[Den Hartog et al.(2003)]{denhartog2003} Den Hartog, E.~A., Lawler, J.~E., Sneden, C., et al.\ 2003, \apjs, 148, 2, 543. doi:10.1086/376940
\bibitem[Den Hartog et al.(2006)]{denhartog2006} Den Hartog, E.~A., Lawler, J.~E., Sneden, C., et al.\ 2006, \apjs, 167, 2, 292. doi:10.1086/508262
\bibitem[Dotter(2016)]{dotter2016} Dotter, A.\ 2016, \apjs, 222, 1, 8. doi:10.3847/0067-0049/222/1/8
\bibitem[Erkal et al.(2016)]{erkal2016} Erkal, D., Sanders, J.~L., \& Belokurov, V.\ 2016, \mnras, 461, 2, 1590. doi:10.1093/mnras/stw1400
\bibitem[Gaia Collaboration et al.(2022)]{gaia2022} Gaia Collaboration, Vallenari, A., Brown, A.G.A., et al. 2022 arXiv e-prints, arXiv:2208.00211
\bibitem[Gaia Collaboration et al.(2021)]{gaiarv21} Gaia Collaboration, Brown, A.~G.~A., Vallenari, A., et al.\ 2021, \aap, 649, A1. doi:10.1051/0004-6361/202039657
\bibitem[Gaia Collaboration et al.(2018)]{gaia2018} Gaia Collaboration, Babusiaux, C., van Leeuwen, F., et al.\ 2018, \aap, 616, A10. doi:10.1051/0004-6361/201832843
\bibitem[Gaia Collaboration(2018)]{gaiarv} Gaia Collaboration\ 2018, VizieR Online Data Catalog, I/345
\bibitem[Gaia Collaboration et al.(2016)]{gaia2016} Gaia Collaboration, Prusti, T., de Bruijne, J.~H.~J., et al.\ 2016, \aap, The Gaia mission, 595, A1. doi:10.1051/0004-6361/201629272
\bibitem[Fitzpatrick et al.(2024)]{fitzpatrick2024} Fitzpatrick, M., Placco, V., Bolton, A., et al.\ 2024, arXiv:2401.01982. doi:10.48550/arXiv.2401.01982
\bibitem[Freeman(2017)]{freeman2017} Freeman, K.~C.\ 2017, \araa, 55, 1. doi:10.1146/annurev-astro-091916-055249
\bibitem[For \& Sneden(2010)]{for2010} For, B.-Q. \& Sneden, C.\ 2010, \aj, 140, 6, 1694. doi:10.1088/0004-6256/140/6/1694
\bibitem[For et al.(2011)]{for2011} For, B.-Q., Sneden, C., \& Preston, G.~W.\ 2011, \apjs, 197, 2, 29. doi:10.1088/0067-0049/197/2/29
\bibitem[Gratton et al.(2019)]{gratton2019} Gratton, R., Bragaglia, A., Carretta, E., et al.\ 2019, \aapr, 27, 1, 8. doi:10.1007/s00159-019-0119-3
\bibitem[Hannaford et al.(1982)]{hannaford1982} Hannaford, P., Lowe, R.~M., Grevesse, N., et al.\ 1982, \apj, 261, 736. doi:10.1086/160384
\bibitem[Hansen et al.(2017)]{hansen2017} Hansen, T.~T., Simon, J.~D., Marshall, J.~L., et al.\ 2017, \apj, 838, 44
\bibitem[Hansen et al.(2020)]{hansen2020gjoll} Hansen, T.~T., Riley, A.~H., Strigari, L.~E., et al.\ 2020, \apj, 901, 23. doi:10.3847/1538-4357/ababa5
\bibitem[Hansen et al.(2018)]{Hansen2018} Hansen, T.~T., Holmbeck, E.~M., Beers, T.~C., et al.\ 2018, \apj, 858, 2, 92. doi:10.3847/1538-4357/aabacc
\bibitem[Hansen et al.(2020)]{hansen2020} Hansen, T.~T., Marshall, J.~L., Simon, J.~D., et al.\ 2020, \apj, 897, 183. doi:10.3847/1538-4357/ab9643
\bibitem[Hansen et al.(2021)]{hansen2021} Hansen, T.~T., Ji, A.~P., Da Costa, G.~S., et al.\ 2021, \apj, 915, 103. doi:10.3847/1538-4357/abfc54
\bibitem[Hansen et al.(2024)]{hansen2024} Hansen, T.~T., Simon, J.~D., Li, T.~S., et al.\ 2024, \apj, 968, 21. doi:10.3847/1538-4357/ad3a52
\bibitem[Harris, C.~R., et al. (2020)]{numpy1} Harris, C.~R., Millman, K.~J., van der Walt, S.~J. et al. \ 2020, \nat, 585, 357-362. doi: 10.1038/s41586-020-2649-2
\bibitem[Hawkins et al.(2023)]{hawkins2023} Hawkins, K., Price-Whelan, A.~M., Sheffield, A.~A., et al.\ 2023, \apj, 948, 2, 123. doi:10.3847/1538-4357/acb698
\bibitem[Helmi(2020)]{helmi2020} Helmi, A.\ 2020, \araa, 58, 205. doi:10.1146/annurev-astro-032620-021917
\bibitem[Holmbeck et al.(2020)]{holmbeck2020} Holmbeck, E.~M., Hansen, T.~T., Beers, T.~C., et al.\ 2020, \apjs, 249, 2, 30. doi:10.3847/1538-4365/ab9c19
\bibitem[Hunter(2007)]{matplotlib} Hunter, J.~D.\ 2007, Computing in Science and Engineering, 9, 90
\bibitem[Malhan \& Ibata(2018)]{ibata2018} Malhan, K. \& Ibata, R.~A.\ 2018, \mnras, 477, 3, 4063. doi:10.1093/mnras/sty912
\bibitem[Ji et al.(2023a)]{ji2023a} Ji, A.~P., Simon, J.~D., Roederer, I.~U., et al.\ 2023, \aj, 165, 100. doi:10.3847/1538-3881/acad84
\bibitem[Ji et al.(2023b)]{ji2023b} Ji, A.~P., Naidu, R.~P., Brauer, K., et al.\ 2023, \mnras, 519, 4467. doi:10.1093/mnras/stac2757
\bibitem[Ji et al.(2020a)]{ji2020a} Ji, A.~P., Li, T.~S., Hansen, T.~T., et al.\ 2020, \aj, 160, 181. doi:10.3847/1538-3881/abacb6
\bibitem[Ji et al.(2020b)]{ji2020b} Ji, A.~P., Li, T.~S., Simon, J.~D., et al.\ 2020, \apj, 889, 27. doi:10.3847/1538-4357/ab6213
\bibitem[Ji et al.(2019a)]{ji2019a} Ji, A.~P., Simon, J.~D., Frebel, A., et al.\ 2019, \apj, 870, 83. doi:10.3847/1538-4357/aaf3bb
\bibitem[Ji et al.(2019b)]{ji2019b} Ji, A.~P., Drout, M.~R., \& Hansen, T.~T.\ 2019, \apj, 882, 1, 40. doi:10.3847/1538-4357/ab3291
\bibitem[Ji et al.(2016)]{ji2016c} Ji, A.~P., Frebel, A., Simon, J.~D., et al.\ 2016, \apj, 830, 93. doi:10.3847/0004-637X/830/2/93
\bibitem[Johnston(1998)]{johnston1998} Johnston, K.~V.\ 1998, \apj, 495, 1, 297. doi:10.1086/305273
\bibitem[Johnston et al.(2008)]{johnston2008} Johnston, K.~V., Bullock, J.~S., Sharma, S., et al.\ 2008, \apj, 689, 2, 936. doi:10.1086/592228
\bibitem[Kelson et al.(2000)]{kelson2000} Kelson, D.~D., Illingworth, G.~D., van Dokkum, P.~G., et al.\ 2000, \apj, 531, 159
\bibitem[Kelson(2003)]{kelson2003} Kelson, D.~D.\ 2003, \pasp, 115, 688. doi:10.1086/375502
\bibitem[Kirby et al.(2023)]{kirby2023} Kirby, E.~N., Ji, A.~P., \& Kovalev, M.\ 2023, \apj, 958, 45. doi:10.3847/1538-4357/acf309
\bibitem[Kirby et al.(2013)]{kirby2013} Kirby, E.~N., Cohen, J.~G., Guhathakurta, P., et al.\ 2013, \apj, 779, 2, 102. doi:10.1088/0004-637X/779/2/102
\bibitem[Saloman \& Kramida(2019)]{kramida2019} Saloman, E.~B. \& Kramida, A.\ 2019, \apjs, 240, 2, 41. doi:10.3847/1538-4365/aaface
\bibitem[Krumholz et al.(2019)]{krumholz2019} Krumholz, M.~R., McKee, C.~F., \& Bland-Hawthorn, J.\ 2019, \araa, 57, 227. doi:10.1146/annurev-astro-091918-104430
\bibitem[Kurucz \& Bell(1995)]{kurucz1995} Kurucz, R.~L. \& Bell, B.\ 1995, Kurucz CD-ROM, Cambridge, MA: Smithsonian Astrophysical Observatory, |c1995, April 15, 1995
\bibitem[Lawler \& Dakin(1989)]{lawler1989} Lawler, J.~E. \& Dakin, J.~T.\ 1989, Journal of the Optical Society of America B Optical Physics, 6, 1457. doi:10.1364/JOSAB.6.001457
\bibitem[Lawler et al.(2001)]{lawler2001a} Lawler, J.~E., Bonvallet, G., \& Sneden, C.\ 2001, \apj, 556, 1, 452. doi:10.1086/321549
\bibitem[Lawler et al.(2001)]{2001b} Lawler, J.~E., Wickliffe, M.~E., den Hartog, E.~A., et al.\ 2001, \apj, 563, 2, 1075. doi:10.1086/323407
\bibitem[Lawler et al.(2004)]{lawler2004} Lawler, J.~E., Sneden, C., \& Cowan, J.~J.\ 2004, \apj, 604, 2, 850. doi:10.1086/382068
\bibitem[Lawler et al.(2006)]{lawler2006} Lawler, J.~E., Den Hartog, E.~A., Sneden, C., et al.\ 2006, \apjs, 162, 1, 227. doi:10.1086/498213
\bibitem[Lawler et al.(2008)]{lawler2008b} Lawler, J.~E., Sneden, C., Cowan, J.~J., et al.\ 2008, \apjs, 178, 1, 71. doi:10.1086/589834
\bibitem[Lawler et al.(2009)]{lawler2009} Lawler, J.~E., Sneden, C., Cowan, J.~J., et al.\ 2009, \apjs, 182, 1, 51. doi:10.1088/0067-0049/182/1/51
\bibitem[Lawler et al.(2011)]{lawler2011} Lawler, J.~E., Bilty, K.~A., \& Den Hartog, E.~A.\ 2011, Journal of Physics B Atomic Molecular Physics, 44, 9, 095001. doi:10.1088/0953-4075/44/9/095001
\bibitem[Lawler et al.(2017)]{lawler2017} Lawler, J.~E., Sneden, C., Nave, G., et al.\ 2017, \apjs, 228, 10. doi:10.3847/1538-4365/228/1/10
\bibitem[Lawler et al.(2014)]{lawler2014} Lawler, J.~E., Wood, M.~P., Den Hartog, E.~A., et al.\ 2014, \apjs, 215, 20. doi:10.1088/0067-0049/215/2/20
\bibitem[Lawler et al.(2013)]{lawler2013} Lawler, J.~E., Guzman, A., Wood, M.~P., et al.\ 2013, \apjs, 205, 11. doi:10.1088/0067-0049/205/2/11
\bibitem[Lawler et al.(2015)]{lawler2015} Lawler, J.~E., Sneden, C., \& Cowan, J.~J.\ 2015, \apjs, 220, 13. doi:10.1088/0067-0049/220/1/13
\bibitem[Letarte et al.(2010)]{letarte2010} Letarte, B., Hill, V., Tolstoy, E., et al.\ 2010, \aap, 523, A17. doi:10.1051/0004-6361/200913413
\bibitem[Lewis et al.(2002)]{lewis2002} Lewis, I.~J., Cannon, R.~D., Taylor, K., et al.\ 2002, \mnras, 333, 279. doi:10.1046/j.1365-8711.2002.05333.x
\bibitem[Li et al.(2022)]{li2022} Li, T.~S., Ji, A.~P., Pace, A.~B., et al.\ 2022, \apj, 928, 30. doi:10.3847/1538-4357/ac46d3
\bibitem[Li et al.(2019)]{li2019} Li, T.~S., Koposov, S.~E., Zucker, D.~B., et al.\ 2019, \mnras, 490, 3508. doi:10.1093/mnras/stz2731
\bibitem[Li et al.(2007)]{li2007} Li, R., Chatelain, R., Holt, R.~A., et al.\ 2007, \physscr, 76, 5, 577. doi:10.1088/0031-8949/76/5/028
\bibitem[Limberg et al.(2021)]{limberg2021} Limberg, G., Santucci, R.~M., Rossi, S., et al.\ 2021, \apjl, 913, 2, L28. doi:10.3847/2041-8213/ac0056
\bibitem[Limberg et al.(2024)]{limberg2024} Limberg, G., Ji, A.~P., Naidu, R.~P., et al.\ 2024, \mnras, 530, 2512. doi:10.1093/mnras/stae969
\bibitem[Ljung et al.(2006)]{ljung2006} Ljung, G., Nilsson, H., Asplund, M., et al.\ 2006, \aap, 456, 3, 1181. doi:10.1051/0004-6361:20065212
\bibitem[McWilliam(1998)]{mcwilliam1998} McWilliam, A.\ 1998, \aj, 115, 1640. doi:10.1086/300289
\bibitem[Malhan et al.(2021)]{malhan2021} Malhan, K., Valluri, M., \& Freese, K.\ 2021, \mnras, 501, 1, 179. doi:10.1093/mnras/staa3597
\bibitem[Malhan \& Ibata(2018)]{streamfinder} Malhan, K. \& Ibata, R.~A.\ 2018, \mnras, 477, 4063. doi:10.1093/mnras/sty912
\bibitem[Marshall et al.(2019)]{marshall2019} Marshall, J.~L., Hansen, T., Simon, J.~D., et al.\ 2019, \apj, 882, 177
\bibitem[Masseron et al.(2014)]{masseron2014} Masseron, T., Plez, B., Van Eck, S., et al.\ 2014, \aap, 571, A47. doi:10.1051/0004-6361/201423956

\bibitem[Mel{\'e}ndez \& Barbuy(2009)]{melendez2009} Mel{\'e}ndez, J. \& Barbuy, B.\ 2009, \aap, 497, 611. doi:10.1051/0004-6361/200811508
\bibitem[Milone et al.(2017)]{milone2017} Milone, A.~P., Piotto, G., Renzini, A., et al.\ 2017, \mnras,464, 3, 3636. doi:10.1093/mnras/stw2531
\bibitem[Mucciarelli et al.(2021)]{mucciarelli2021} Mucciarelli, A., Bellazzini, M., \& Massari, D.\ 2021, \aap, 653, A90. doi:10.1051/0004-6361/202140979
\bibitem[Norris et al.(2017)]{norris2017} Norris, J.~E., Yong, D., Venn, K.~A., et al.\ 2017, \apjs, 230, 28. doi:10.3847/1538-4365/aa755e
\bibitem[O'Brian et al.(1991)]{obrian1991} O'Brian, T.~R., Wickliffe, M.~E., Lawler, J.~E., et al.\ 1991, Journal of the Optical Society of America B Optical Physics, 8, 1185. doi:10.1364/JOSAB.8.001185
\bibitem[Placco et al.(2021)]{placco2021} Placco, V.~M., Sneden, C., Roederer, I.~U., et al.\ 2021, Research Notes of the American Astronomical Society, 5, 92. doi:10.3847/2515-5172/abf651
\bibitem[Placco et al.(2014)]{placco2014} Placco, V.~M., Frebel, A., Beers, T.~C., et al.\ 2014, \apj, 797, 21. doi:10.1088/0004-637X/797/1/21
\bibitem[Roederer et al.(2022)]{roederer2022} Roederer, I.~U., Lawler, J.~E., Den Hartog, E.~A., et al.\ 2022, \apjs, 260, 2, 27. doi:10.3847/1538-4365/ac5cbc
\bibitem[Roederer \& Lawler(2012)]{roederer2012} Roederer, I.~U. \& Lawler, J.~E.\ 2012, \apj, 750, 76. doi:10.1088/0004-637X/750/1/76
\bibitem[Roederer et al.(2014)]{roederer2014} Roederer, I.~U., Preston, G.~W., Thompson, I.~B., et al.\ 2014, \aj, 147, 136
\bibitem[Roederer(2011)]{roederer2011} Roederer, I.~U.\ 2011, \apjl, 732, 1, L17. doi:10.1088/2041-8205/732/1/L17
\bibitem[Ruffoni et al.(2014)]{ruffoni2014} Ruffoni, M.~P., Den Hartog, E.~A., Lawler, J.~E., et al.\ 2014, \mnras, 441, 3127. doi:10.1093/mnras/stu780
\bibitem[Ryabchikova et al.(2015)]{ryabchikova2015} Ryabchikova, T., Piskunov, N., Kurucz, R.~L., et al.\ 2015, \physscr, 90, 5, 054005. doi:10.1088/0031-8949/90/5/054005
\bibitem[Schlafly \& Finkbeiner(2011)]{schlafly2011} Schlafly, E.~F. \& Finkbeiner, D.~P.\ 2011, \apj, 737, 103. doi:10.1088/0004-637X/737/2/103
\bibitem[Sharp et al.(2006)]{sharp2006} Sharp, R., Saunders, W., Smith, G., et al.\ 2006, \procspie, 6269, 62690G. doi:10.1117/12.671022
\bibitem[Shipp et al.(2023)]{shipp2023} Shipp, N., Panithanpaisal, N., Necib, L., et al.\ 2023, \apj, 949, 44. doi:10.3847/1538-4357/acc582

\bibitem[Shipp et al.(2019)]{shipp2019} Shipp, N., Li, T.~S., Pace, A.~B., et al.\ 2019, \apj, 885, 3. doi:10.3847/1538-4357/ab44bf
\bibitem[Shipp et al.(2018)]{shipp2018} Shipp, N., Drlica-Wagner, A., Balbinot, E., et al.\ 2018, \apj, 862, 114. doi:10.3847/1538-4357/aacdab
\bibitem[Shetrone et al.(2003)]{shetrone2003} Shetrone, M., Venn, K.~A., Tolstoy, E., et al.\ 2003, \aj, 125, 684. doi:10.1086/345966
\bibitem[Sitnova et al.(2024)]{sitnova2024} Sitnova, T.~M., Yuan, Z., Matsuno, T., et al.\ 2024, \aap, 690, A331. doi:10.1051/0004-6361/202450981
\bibitem[Siqueira Mello et al.(2014)]{SiqueraMello2014} Siqueira Mello, C., Hill, V., Barbuy, B., et al.\ 2014, \aap, 565, A93. doi:10.1051/0004-6361/201423826
\bibitem[Sneden(1973)]{sneden1973} Sneden, C.~A.\ 1973, Ph.D. Thesis, THE UNIVERSITY OF TEXAS AT AUSTIN
\bibitem[Sneden et al.(1997)]{sneden1997} Sneden, C., Kraft, R.~P., Shetrone, M.~D., et al.\ 1997, \aj, 114, 1964. doi:10.1086/118618
\bibitem[Sneden et al.(2009)]{sneden2009} Sneden, C., Lawler, J.~E., Cowan, J.~J., et al.\ 2009, \apjs, 182, 1, 80. doi:10.1088/0067-0049/182/1/80
\bibitem[Sneden et al.(2008)]{sneden2008} Sneden, C., Cowan, J.~J., \& Gallino, R.\ 2008, \araa, 46, 241. doi:10.1146/annurev.astro.46.060407.145207
\bibitem[Sneden et al.(2014)]{sneden2014} Sneden, C., Lucatello, S., Ram, R.~S., et al.\ 2014, \apjs, 214, 2, 26. doi:10.1088/0067-0049/214/2/26
\bibitem[Sobeck et al.(2007)]{sobeck2007} Sobeck, J.~S., Lawler, J.~E., \& Sneden, C.\ 2007, \apj, 667, 1267. doi:10.1086/519987
\bibitem[Sobeck et al.(2011)]{sobeck2011} Sobeck, J.~S., Kraft, R.~P., Sneden, C., et al.\ 2011, \aj, 141, 175
\bibitem[Tavangar et al.(2022)]{tavangar2022} Tavangar, K., Ferguson, P., Shipp, N., et al.\ 2022, \apj, 925, 118. doi:10.3847/1538-4357/ac399b
\bibitem[Tody(1986)]{tody1986} Tody, D.\ 1986, \procspie, 733
\bibitem[Tody(1993)]{tody1993} Tody, D.\ 1993, Astronomical Data Analysis Software and Systems II, 173
\bibitem[Tolstoy et al.(2009)]{tolstoy2009} Tolstoy, E., Hill, V., \& Tosi, M.\ 2009, \araa, 47, 371. doi:10.1146/annurev-astro-082708-101650
\bibitem[Travaglio et al.(2004)]{Travaglio2004} Travaglio, C., Gallino, R., Arnone, E., et al.\ 2004, \apj, 601, 2, 864. doi:10.1086/380507
\bibitem[Usman et al.(2024)]{usman2024} Usman, S.~A., Ji, A.~P., Li, T.~S., et al.\ 2024, \mnras, 529, 3, 2413. doi:10.1093/mnras/stae185
\bibitem[van der Walt et al.(2011)]{numpy} van der Walt, S., Colbert, S.~C., \& Varoquaux, G.\ 2011, Computing in Science and Engineering, 13, 22
\bibitem[Venn et al.(2012)]{venn2012} Venn, K.~A., Shetrone, M.~D., Irwin, M.~J., et al.\ 2012, \apj, 751, 102. doi:10.1088/0004-637X/751/2/102
\bibitem[Wan et al.(2020)]{wan2020} Wan, Z., Lewis, G.~F., Li, T.~S., et al.\ 2020, \nat, 583, 768. doi:10.1038/s41586-020-2483-6
\bibitem[Wenger et al.(2000)]{wenger2000} Wenger, M., Ochsenbein, F., Egret, D., et al.\ 2000, \aaps, 143, 9. doi:10.1051/aas:2000332
\bibitem[Wood et al.(2013)]{wood2013} Wood, M.~P., Lawler, J.~E., Sneden, C., et al.\ 2013, \apjs, 208, 27. doi:10.1088/0067-0049/208/2/27
\bibitem[Wood et al.(2014a)]{wood2014a} Wood, M.~P., Lawler, J.~E., Den Hartog, E.~A., et al.\ 2014, \apjs, 214, 18. doi:10.1088/0067-0049/214/2/18
\bibitem[Wood et al.(2014b)]{wood2014b} Wood, M.~P., Lawler, J.~E., Sneden, C., et al.\ 2014, \apjs, 211, 20. doi:10.1088/0067-0049/211/2/20
\bibitem[Xylakis-Dornbusch et al.(2024)]{xylakis-dornbusch2024} Xylakis-Dornbusch, T., Hansen, T.~T., Beers, T.~C., et al.\ 2024, \aap, 688, A123. doi:10.1051/0004-6361/202449376
\bibitem[Yu \& Derevianko(2018)]{yu2018} Yu, Y. \& Derevianko, A.\ 2018, Atomic Data and Nuclear Data Tables, 119, 263. doi:10.1016/j.adt.2017.03.002
\bibitem[Zevin et al.(2019)]{zevin2019} Zevin, M., Kremer, K., Siegel, D.~M., et al.\ 2019, \apj, 886, 1, 4. doi:10.3847/1538-4357/ab498b

\end{thebibliography}
\end{document}